\documentclass[12pt]{article}
\usepackage{epsfig}
\usepackage{amssymb}
\usepackage{amsmath}
\usepackage{amsfonts}
\usepackage{graphicx}
\usepackage{mathrsfs}
\usepackage[dvips]{color}
\usepackage{multirow}

\newcommand{\bsigma}{\boldsymbol{\sigma}}

\newcommand{\R}{\mathbb{R}}
\newcommand{\C}{\mathbb{C}}
\newcommand{\Z}{\mathbb{Z}}

\newcommand{\ff}{\mathfrak{f}}
\newcommand{\fg}{\mathfrak{g}}

\newcommand{\fz}{\mathfrak{z}}

\newcommand{\fK}{\mathfrak{K}}
\newcommand{\bfK}{\boldsymbol{\mathfrak{K}}}

\newcommand{\fM}{\mathfrak{M}}

\newcommand{\bfe}{\mathbf{e}}

\newcommand{\bk}{\mathbf{k}}

\newcommand{\bfr}{\mathbf{r}}

\newcommand{\bu}{\mathbf{u}}

\newcommand{\bH}{\mathbf{H}}
\newcommand{\bI}{\mathbf{I}}

\newcommand{\bM}{\mathbf{M}}

\newcommand{\bS}{\mathbf{S}}

\newcommand{\bU}{\mathbf{U}}

\newcommand{\cH}{\mathcal{H}}

\newcommand{\cF}{\mathcal{F}}

\newcommand{\cK}{\mathcal{K}}

\newcommand{\cS}{\mathcal{S}}

\newcommand{\cU}{\mathcal{U}}

\newcommand{\be}{\begin{equation}}
\newcommand{\ee}{\end{equation}}
\newcommand{\bea}{\begin{eqnarray}}
\newcommand{\eea}{\end{eqnarray}}
\newcommand{\nn}{\nonumber}

\newcommand{\ed}{\end{document}}

\newcommand{\bi}{\begin{itemize}}
\newcommand{\ei}{\end{itemize}}

\newcommand{\bce}{\begin{center}}
\newcommand{\ece}{\end{center}}
\newcommand{\sA}{\mathscr{A}}
\newcommand{\sB}{\mathscr{B}}

\newcommand{\sD}{\mathscr{D}}

\newcommand{\sF}{\mathscr{F}}
\newcommand{\sG}{\mathscr{G}}

\newcommand{\sR}{\mathscr{R}}

\newcommand{\sT}{\mathscr{T}}
\newcommand{\sV}{\mathscr{V}}

\newcommand{\RE}{{\rm Re}}
\newcommand{\IM}{{\rm Im}}

\newcommand{\bPsi}{{\boldsymbol{\Psi}}}
\newcommand{\bPhi}{{\boldsymbol{\Phi}}}
\newcommand{\bPi}{{\boldsymbol{\Pi}}}
\newcommand{\bcK}{{\boldsymbol{\cK}}}

\newcommand{\bfM}{{\boldsymbol{\fM}}}
\newcommand{\bcH}{{\boldsymbol{\cH}}}
\newcommand{\bcU}{{\boldsymbol{\cU}}}

\newcommand{\bzero}{{\boldsymbol{0}}}

\newcommand{\for}{{\mbox{\rm for}}}

\begin{document}

\title{Fundamental transfer matrix and dynamical formulation of stationary scattering in two and three dimensions}


\author{Farhang Loran\thanks{E-mail address: loran@iut.ac.ir}~ and
Ali~Mostafazadeh\thanks{E-mail address:
amostafazadeh@ku.edu.tr}\\[6pt]
$^{*}$Department of Physics, Isfahan University of Technology, \\ Isfahan 84156-83111, Iran\\[6pt]
$^\dagger$Departments of Mathematics and Physics, Ko\c{c}
University,\\  34450 Sar{\i}yer, Istanbul, Turkey}

\date{ }
\maketitle

\begin{abstract}

We offer a consistent dynamical formulation of stationary scattering in two and three dimensions that is based on a suitable multidimensional generalization of the transfer matrix. This is a linear operator acting in an infinite-dimensional function space which we can represent as a $2\times 2$ matrix with operator entries. This operator encodes the information about the scattering properties of the potential and enjoys an analog of the composition property of its one-dimensional ancestor. Our results improve an earlier attempt in this direction [Phys.\ Rev.~A \textbf{93}, 042707 (2016)] by elucidating the role of the evanescent waves. In particular, we show that a proper formulation of this approach requires the introduction of a pair of intertwined transfer matrices each related to the time-evolution operator for an effective non-unitary quantum system. We study the application of our findings in the treatment of the scattering problem for delta-function potentials in two and three dimensions and clarify its implicit regularization property which circumvents the singular terms appearing in the standard treatments of these potentials. We also discuss the utility of our approach in characterizing invisible (scattering-free) potentials and potentials for which the first Born approximation provides the exact expression for the scattering amplitude.
\vspace{2mm}


\end{abstract}

\section{Introduction}

Scattering of waves is a natural phenomenon of great importance. This was recognized quite early and has led to a comprehensive study of this phenomenon in the nineteenth and twentieth centuries. The formulation of quantum scattering theory \cite{Born-1926} by Born in 1926, the introduction of the concept of the scattering (S) matrix \cite{wheeler-1937} by Wheeler in 1937, and the developments of the Green's function methods for treating scattering problems \cite{Lippmann-Schwinger} by Lippmann and Schwinger in 1950 are the milestones of our present understanding of the subject. The study and applications of the Born series, the S-matrix, and the Lippman-Schwinger equation constitute an integral part of the standard textbook treatments of scattering theory \cite{sakurai,Newton-ST}. They have also provided the main impetus for developing rigorous mathematical theories of scattering \cite{reed-simon3,lax,yafaev}. There is essentially no conceptual difference between the utility of these concepts and theories in dealing with the scattering of waves in different dimensions. However, in one dimension, one has the option of employing an alternative tool, called the transfer matrix \cite{jones-1941,abeles,thompson}. This is an object, which similarly to the S-matrix, stores the information about the scattering features of the system and enjoys a composition property that makes it an ideal tool for dealing with multilayer and locally periodic systems \cite{yeh,griffiths,yeh-book,tjp-2020}.

Consider a short-range potential in one dimension, $v:\R\to\C$, so that $|v(x)|$ tends to zero faster than $|x|^{-1}$ as $x\to\pm\infty$. Then, every solution of the stationary Schr\"odinger equation,
    \be
    -\psi''(x)+v(x)\psi(x)=k^2\psi(x)~~~~x\in\R,
    \label{sh-eq-1D}
    \ee
satisfies
    \[\psi(x)\to \left\{\begin{array}{ccc}
    A_-e^{ikx}+B_- e^{-ikx}~~~\for~~~x\to-\infty,\\
    A_+ e^{ikx}+B_+ e^{-ikx}~~~\for~~~x\to+\infty,\end{array}\right.\]
where $k\in\R^+$ is a wavenumber, and $A_\pm$ and $B_\pm$ are complex coefficients. The transfer matrix of the potential $v$ is a $2\times 2$ matrix $\bM$ that relates $A_\pm$ and $B_\pm$ according to
    \be
    \left[\begin{array}{c}
    A_+\\
    B_+\end{array}\right]=\bM
    \left[\begin{array}{c}
    A_-\\
    B_-\end{array}\right].
    \label{M-def-1D}
    \ee
This equation determines $\bM$ in a unique manner provided that it is independent of $A_-$ and $B_-$, \cite{epjp-2019}. We can also identify the S-matrix for $v$ by the $2\times 2$ matrix $\bS$ that satisfies \cite{muga-review,bookchapter},
    \be
    \left[\begin{array}{c}
    A_+\\
    B_-\end{array}\right]=\bS
    \left[\begin{array}{c}
    A_-\\
    B_+\end{array}\right].
    \label{S-def-1D}
    \ee

If we enforce (\ref{M-def-1D}) and (\ref{S-def-1D}) for the coefficients $A_\pm$ and $B_\pm$ of the left- and right-incident scattering solutions $\psi_{l/r}$ of (\ref{sh-eq-1D}), i.e., those fulfilling
    \begin{align}
    &\psi_l(x)\to\left\{\begin{array}{ccc}
    e^{ikx}+R^l e^{-ikx} &\for&x\to-\infty,\\
    T^l\,e^{ikx}&\for&x\to+\infty,\end{array}\right.
    &&\psi_r(x)\to\left\{\begin{array}{ccc}
    T^r\,e^{-ikx}&\for&x\to-\infty,\\
    e^{-ikx}+R^r e^{ikx} &\for&x\to+\infty,\end{array}\right.
    \end{align}
we can express the left/right reflection and transmission amplitudes, $R^{l/r}$ and $T^{l/r}$, of the potential in terms of the entries of $\bM$ and $\bS$, \cite{tjp-2020,muga-review,bookchapter};
    \begin{align}
    &R^l=-\frac{M_{21}}{M_{22}},
    &&R^r=\frac{M_{12}}{M_{22}},
    &&T^{l/r}=\frac{1}{M_{22}},
    \label{RRT-M}\\
    &R^l=S_{21},
    &&R^r=S_{12},
    &&T^{l/r}=S_{11}=S_{22}.
    \label{RRT-S}
    \end{align}
These relations show that both $\bM$ and $\bS$ contain the information about the scattering properties of the potential. Therefore, their determination is equivalent to the solution of the scattering problem.

The advantage of the transfer matrix over the S-matrix is its composition property. If we divide $\R$ into $n$ adjacent intervals of the form,
    \begin{align*}
    &I_1:=(-\infty,a_1),
    &&I_2:=[a_1,a_2),
    &&I_3:=[a_2,a_3),
    &&\cdots
    &&I_{n-1}:=[a_{n-2},a_{n-1}),
    &&I_n:=[a_{n-1},\infty),
    \end{align*}
where $a_1,a_2,\cdots,a_{n-1}\in\R$ such that $a_1<a_2<\cdots<a_{n-1}$, let $v_j:\R\to\C$ be the truncation of $v$ given by
    \[v_j(x):=\left\{\begin{array}{ccc}
    v(x) & \for &x\in I_j,\\
    0 & \for &x\notin I_j,\end{array}\right.\]
and $\bM_j$ be the transfer matrix of $v_j$, then the following composition rule holds \cite{tjp-2020}.
    \be
    \bM=\bM_n\bM_{n-1}\cdots\bM_1.
    \label{compose-1D}
    \ee
This relation is the main reason for the wide range of applications of the transfer matrix \cite{yeh,teitler-1970,berreman-1972,abrahams-1980,ardos-1982,pendry-1982,levesque,sheng-1996,schubert-1996,wang-2001,wortmann-2002,katsidis-2002,Hao-2008,li-2009,zhan-2013}. It has provided the main guidline for the development of the multichannel \cite{pereyray-1998a,pereyray-2002,pereyray-2005,Shukla-2005,anzaldo-meneses-2007}
and higher-dimensional
\cite{pendry-1984,pendry-1990a, pendry-1990b,pendry-1994,mclean,ward-1996,pendry-1996} generalizations of the transfer matrix~(\ref{M-def-1D}). The latter generalizations involve discretization of the configuration or momentum space variables along the normal direction(s) to the scattering/propagation axis of the wave and yield large numerical transfer matrices whose treatment requires appropriate numerical schemes. Therefore, unlike the S-matrix, they do not arise from a fundamental concept with a universal definition.

Ref.~\cite{pra-2016} pursues a different route and arrives at a higher-dimensional notion of transfer matrix whose definition does not involve any discretization or approximation scheme. This is a linear operator acting in an infinite-dimensional function space. It admits a suitable realization in terms of a $2\times 2$ matrix with operator entries that has the same structure as the transfer matrix in one dimension. It also shares the basic properties of the latter, namely it carries the information about the scattering features of the potential and possesses a similar composition property. Another appealing feature of this transfer matrix is that it admits a Dyson series expansion. This follows from a higher-dimensional generalization of a dynamical formulation of stationary scattering in one dimension \cite{ap-2014} where the standard transfer matrix is identified with the S-matrix of an effective non-unitary two-level quantum system \cite{pra-2014a}. This observation has interesting applications in constructing tunable unidirectionally invisible potentials \cite{tjp-2020}, extending the notion of the transfer matrix to long-range potentials \cite{jpa-2020b},  and performing low-energy scattering calculations in one dimension \cite{jmp-2021}. See also \cite{jpa-2021}.

The application of the transfer matrix of Ref.~\cite{pra-2016} to delta-function potentials in two and three dimensions turns out not to involve the unwanted singularities of the standard treatments of these potentials \cite{mead,manuel,Adhikari1,Adhikari2,Mitra,Rajeev-1999,Nyeo,Camblong,ap-2019}. Furthermore, this approach to scattering theory opens up a new line of attack on the problem of identifying invisible (scattering-free) complex potentials \cite{prsa-2016,ol-2017,pra-2017,pra-2019} and admits a useful electromagnetic counterpart \cite{jpa-2020}.

A careful examination of the analysis of Ref.~\cite{pra-2016} reveals certain underlying assumptions and implicit rules of calculation that require further justification. These shortcomings may be traced back to an inadequate treatment of the role of the evanescent waves. These waves do not enter the general expression for the transfer matrix in an explicit manner, but their contribution to the scattering phenomenon cannot be ignored in general. In particular, they play an important role in the identification of the correct form of the composition property of the transfer matrix which makes it into a powerful practical tool. The purpose of the present article is to offer a comprehensive refinement of the concept of a fundamental transfer matrix in two and three dimensions that eliminates the shortcomings of Ref.~\cite{pra-2016} and yields a consistent alternative to the standard treatment of stationary scattering.

The organization of this article is as follows. In Sec.~2 we introduce a variant of the transfer matrix of Ref.~\cite{pra-2016} in two dimensions, called the fundamental transfer matrix, which takes into account the contribution of the evanescent waves. In Sec.~3, we identify a Hamiltonian operator whose evolution operator yields the fundamental transfer matrix. In Sec.~4, we derive the composition rule for the fundamental transfer matrix. This turns out to require the introduction of an alternative (auxiliary) transfer matrix which is also related to the evolution operator of an effective quantum system. In Sec.~5, we examine the utility of the auxiliary transfer matrix in the solution of the scattering problem and derive a basic formula describing its relationship to the fundamental transfer matrix.  In Sec.~6, we present the application of our general results to delta-function potentials in two dimensions. Here we elucidate the conceptual basis for the implicit regularization property of the fundamental transfer matrix that makes it avoid the singularities of the standard treatments of these potentials. In  Sec.~7, we discuss the application of the fundamental transfer matrix in constructing potentials that display perfect broadband omnidirectional invisibility in two dimensions. Here we also present a characterization theorem for potentials for which the first Born approximation gives the exact expression for the scattering amplitude. In Sec.~8, we generalize the results of Secs. 2-7 to three dimensions. Sec.~9 provides a summary of our findings, and appendices include the derivation of some of the results we present in the text.

\section{Fundamental transfer matrix in 2D}
\label{Sec2}

Consider the stationary Schr\"odinger equation in two dimensions,
    \be
    [-\partial_x^2-\partial_y^2+v(x,y)]\psi(x,y)=k^2\psi(x,y),~~~~~~(x,y)\in\R^2,
    \label{sch-eq}
    \ee
for a short-range potential $v:\R^2\to\C$ and a wavenumber $k\in\R^+$. This equation has a unique solution possessing the following asymptotic expression \cite{yafaev}.
    \be
    \psi(\bfr)=
    \frac{1}{2\pi}\Big[e^{i\bk_0\cdot\bfr}+\sqrt{\frac{i}{kr}}\,e^{ikr}\ff(\theta) \Big]+o(r^{-1/2})
    ~~\for~~r\to\infty,
    \label{scattering}
    \ee
where $\bk_0\in\R^2$ is the incident wave vector, $\bfr:=x\, \bfe_x+y\,\bfe_y$, $\bfe_j$ is the unit vector along the $j$-axis for $j\in\{x,y\}$, $(r,\theta)$ are the polar coordinates of $\bfr$, $\bk_0\cdot\bfr:=kr\cos(\theta-\theta_0)$, $\theta_0$ is the incidence angle which determines $\bk_0/k$ according to $\bk_0/k=\cos\theta_0\,\bfe_x+\sin\theta_0\,\bfe_y$,
$\ff$ is a complex-valued function called the scattering amplitude, and $o(r^{-1/2})$ stands for a function of $r$ and $\theta$ such that
    \[\lim_{r\to\infty}r^{1/2}o(r^{-1/2})=0.\]
By solving the scattering problem for the potential $v$, we mean determining $\ff(\theta)$ for every wavenumber $k$ and incidence angle $\theta_0$.

\begin{figure}
    \begin{center}
        \includegraphics[scale=.45]{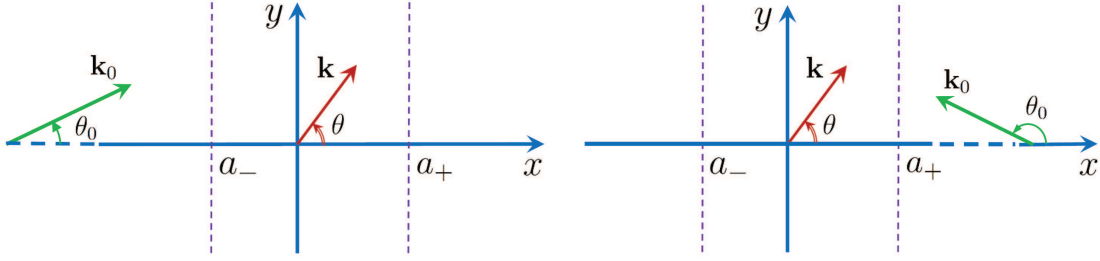}
        \caption{Schematic view of the scattering setup for a
        left-incident wave (on the left) and a right-incident
        wave (on the right). $\bk_0$ and $\bk$ are respectively the
        incident and scattered wave vectors. For the left- and
        right-incident waves the incidence angle $
        \theta_0$ takes values in $(-\frac{\pi}{2},\frac{
        \pi}{2})$ and $(\frac{\pi}{2},\frac{3\pi}{2})$.
        Dashed lines represent $x=a_\pm$.}
        \label{fig1}
    \end{center}
    \end{figure}

Let us adopt a coordinate system in which the source of the incident
wave and the detectors used to observe the scattered wave lie on the
planes $x=\pm\infty$. If the source of the incident wave lies at
$x=-\infty$, $\theta_0\in(\mbox{$-\frac{\pi}{2},\frac{\pi}{2}$})$,
and we speak of a left-incident wave. If the source lies at
$x=+\infty$, $\theta_0\in(\mbox{$\frac{\pi}{2},\frac{3\pi}{2}$})$,
and we have a right-incident wave. See~Fig.~\ref{fig1}.
We use $\ff^l$ and $\ff^r$ to label the scattering amplitudes for
the left- and right-incident waves, respectively;
    \be
    \ff(\theta)=\left\{\begin{array}{ccc}
    \ff^l(\theta)&\for&\theta_0\in(\mbox{$-\frac{\pi}{2},\frac{\pi}{2}$}),\\[6pt]
    \ff^r(\theta)&\for&\theta_0\in(\mbox{$\frac{\pi}{2},\frac{3\pi}{2}$}).
    \end{array}\right.
    \label{ff-f-def}
    \ee

Let $\cF_{y,p}$ and $\cF^{-1}_{p,y}$ denote the Fourier transformation of a function of $y$ and its inverse, i.e.,
    \bea
    \cF_{y,p}\{f(y)\}&:=&\tilde f(p):=\int_{-\infty}^\infty dy\,e^{-ip y} f(y),
    \label{Fourier}\\
    \cF^{-1}_{p,y}\{g(p)\}&:=&\frac{1}{2\pi}\int_{-\infty}^\infty dp\,e^{ip y} g(p),
    \label{inv-Fourier}
    \eea
where $f,g:\R\to\C$ are functions.\footnote{Throughout this article we use the term ``function'' to mean a classical function or a tempered distribution \cite{strichartz}.} Performing the Fourier transform of both sides of (\ref{sch-eq}) with respect to $y$, we find
    \be
    -\tilde\psi''(x,p)+(\widehat\sV(x)\tilde\psi)(x,p)
    =\varpi(p)^2\,\tilde\psi(x,p),\quad\quad\quad (x,p)\in\R^2,
    \label{sch-eq-FT}
    \ee
where a prime stands for differentiation with respect to $x$, $\tilde \psi(x,p):=\cF_{y,p}\{\psi(x,y)\}$,
    \be
    (\widehat\sV(x)\tilde f)(p):=\cF_{y,p}\{v(x,y)f(y)\}=
    \frac{1}{2\pi}\int_{-\infty}^\infty\!\!dq~\tilde v(x,p-q)\tilde f(q),
    \label{v-def}
    \ee
and
    \bea
    \varpi(p)&:=&\left\{\begin{array}{ccc}
    \sqrt{k^2-p^2} & \for & |p|< k,\\
    i\sqrt{p^2-k^2} & \for & |p|\geq k.\end{array}\right.
    \label{varpi}
    \eea

An important feature of potential scattering in two (and higher) dimensions, which has no counterpart in one dimension, is the presence of evanescent waves. To elucidate their role, in the following, we confine our attention to the class of potentials whose supports lie between a pair of lines that are parallel to the $y$-axis. In other words, we suppose that there are real numbers $a_\pm$ such that $a_-<a_+$ and
    \be
    v(x,y)=0~~\for~~x\notin[a_-,a_+].
    \label{supp-x}
    \ee
Then, $\tilde v(x,p)=0$ for $x\notin[a_-,a_+]$, and  (\ref{sch-eq-FT}) gives
    \[[\partial_x^2+\varpi(p)^2]\tilde\psi(x,p)=0~~\for~~x\notin[a_-,a_+].\]
Solving this equation and performing the inverse Fourier transform with respect to $p$, we can write $\psi$ in the form
    \be
    \psi=\psi_{\rm os}+\psi_{\rm ev},
    \label{decompose}
    \ee
where $\psi_{\rm os},\psi_{\rm ev}:\R^2\to\C$ are respectively functions representing the oscillating and evanescent waves outside the region defined by $a_-<x<a_+$ in $\R^2$. This means that
    \bea
    \psi_{\rm os}(x,y)&=&\frac{1}{2\pi}
    \int_{-k}^k dp\;\left[A_\pm(p) e^{i\varpi(p)x}+B_\pm(p) e^{-i\varpi(p)x}\right]e^{ip y}~~\for~~\pm x\geq \pm a_\pm,
    \label{psi-o1}\\
    \psi_{\rm ev}(x,y)&=&\frac{1}{2\pi}
    \int_{|p|\geq k}\!\!\!dp\: C_\pm(p) e^{\mp|\varpi(p)|x} e^{ipy}~~\for~~\pm x\geq \pm a_\pm,
    \label{psi-e1}
    \eea
where $A_\pm,B_\pm,C_\pm:\R\to\C$ are functions such that\footnote{More precisely, $A_\pm$ and $B_\pm$ (respectively $C_\pm$) are tempered distributions supported in the open interval $(-k,k)$ (respectively $\R\setminus (-k,k)$.)}
    \begin{align}
    &A_\pm(p)=B_\pm(p)=0~~\for~~|p|\geq k,
    &&C_\pm(p)=0~~\for~~|p|<k.
    \label{ABC-bound}
    \end{align}
In the following we use $\sF_k$ to denote the vector space of functions $\phi$ such that $\phi(p)=0$ for $|p|\geq k$.
Then (\ref{ABC-bound}) states that $A_\pm, B_\pm\in\sF_k$.

We also introduce,
    \begin{align}
    &\sB_-(p):=B_-(p)+C_-(p)=\left\{\begin{array}{ccc}
    B_-(p) &\for & |p|<k,\\
    C_-(p) &\for & |p|\geq k,\end{array}\right.
    \label{sB=}\\
    &\sA_+(p):=A_+(p)+C_+(p)=\left\{\begin{array}{ccc}
    A_+(p) &\for & |p|<k,\\
    C_+(p) &\for & |p|\geq k,\end{array}\right.
    \label{sA=}
    \end{align}
for future use, and employ (\ref{Fourier}) and (\ref{decompose}) -- (\ref{sA=}) to conclude that
    \bea
    \tilde\psi(x,p)&=&\left\{
    \begin{array}{ccc}
    A_-(p) e^{i\varpi(p)x}+\sB_-(p) e^{-i\varpi(p)x} & \for & x\leq a_-,\\[6pt]
    \sA_+(p) e^{i\varpi(p)x}+B_+(p) e^{-i\varpi(p)x} & \for & x\geq a_+.
    \end{array}\right.
    \label{asym-1}
    \eea

According to (\ref{psi-e1}), $\psi_{\rm ev}(x,y)\to 0$ as $x\to\pm\infty$. In light of (\ref{scattering}) and (\ref{decompose}), this means that the scattering properties of the potential are encoded in the functions $A_\pm$ and $B_\pm$. To see this, we examine the scattering of the left- and right-incident waves separately, use $A^{l/r}_\pm$ and $B^{l/r}_\pm$ to identify the functions $A_\pm$ and $B_\pm$ associated with  left-/right-incident waves, and set
    \begin{align}
    &p:=k\sin\theta,
    &&p_0:=k\sin\theta_0.
    \label{pp}
    \end{align}
For a left-indecent wave, $\theta_0\in (-\frac{\pi}{2},\frac{\pi}{2})$, and as we show in \cite[Appendix~A]{pra-2016},
    \begin{align}
    &A^l_-(p)=2\pi\delta(p-p_0)=\frac{2\pi\,\delta(\sin\theta-\sin\theta_0)}{k}=
    \frac{2\pi\,\delta(\theta-\theta_0)}{k|\cos\theta_0|},
    \quad\quad\quad B^l_+(p)=0,
    \label{z81}\\[6pt]
    &\ff^l(\theta)=-\frac{i}{\sqrt{2\pi}}\times \left\{
    \begin{array}{ccc}
    k |\cos\theta| A^l_+(k \sin\theta)-2\pi\delta(\theta-\theta_0) &\for & \theta\in(-\frac{\pi}{2},\frac{\pi}{2}),\\[6pt]
    k |\cos\theta| B^l_-(k \sin\theta) &\for & \theta\in(\frac{\pi}{2},\frac{3\pi}{2}).\end{array}\right.
    \label{z82}
    \end{align}
Similarly, for a right-incident wave, where $\theta_0\in (\frac{\pi}{2},\frac{3\pi}{2})$, we have \cite{prsa-2016},
    \begin{align}
    &A^r_-(p)=0,\quad\quad\quad\quad
    B^r_+(p)=2\pi\delta(p-p_0)=\frac{2\pi\,\delta(\theta-\theta_0)}{k|\cos\theta_0|},
    \label{z83}\\[6pt]
    &\ff^r(\theta)=-\frac{i}{\sqrt{2\pi}}\times \left\{
    \begin{array}{ccc}
    k |\cos\theta| A^r_+(k \sin\theta) &\for & \theta\in(-\frac{\pi}{2},\frac{\pi}{2}),\\[6pt]
    k |\cos\theta| B^r_-(k \sin\theta)-2\pi\delta(\theta-\theta_0) &\for & \theta\in(\frac{\pi}{2},\frac{3\pi}{2}).\end{array}\right.
    \label{z84}
    \end{align}
In view of Eqs.~(\ref{z82}) and (\ref{z84}) and the fact that $k|\cos\theta|=\varpi(p)$, we can express the scattering amplitudes $\ff^{l/r}$ in terms of the dimensionless coefficient functions,
    \begin{align}
    &\breve A_\pm:=\varpi\,A_\pm,
    &&\breve B_\pm:=\varpi\,B_\pm,
    \label{breves}
    \end{align}
which also belong to $\sF_k$. This gives
\begin{align}
    &\breve B^l_+(p)=\breve A^r_-(p)=0,\quad\quad\quad\quad
    \breve A^l_-(p)=\breve B^r_+(p)=2\pi\varpi(p_0)\,\delta(p-p_0),
    \label{z81b}\\[6pt]
    &\ff^l(\theta)=-\frac{i}{\sqrt{2\pi}}\times\left\{
    \begin{array}{ccc}
    \breve A^l_+(k \sin\theta)-2\pi \delta(\theta-\theta_0) &\for & \theta\in(-\frac{\pi}{2},\frac{\pi}{2}),\\[6pt]
    \breve B^l_-(k \sin\theta) &\for & \theta\in(\frac{\pi}{2},\frac{3\pi}{2}),\end{array}\right.
    \label{z82b}\\[6pt]
    &\ff^r(\theta)=-\frac{i}{\sqrt{2\pi}}\times\left\{
    \begin{array}{ccc}
    \breve A^r_+(k \sin\theta) &\for & \theta\in(-\frac{\pi}{2},\frac{\pi}{2}),\\[6pt]
    \breve B^r_-(k \sin\theta)-2\pi\delta(\theta-\theta_0)  &\for & \theta\in(\frac{\pi}{2},\frac{3\pi}{2}).\end{array}\right.
    \label{z84b}
    \end{align}
Notice that $\ff^{l/r}(\theta)$ for $\theta\in(-\frac{\pi}{2},\frac{\pi}{2})$ and $\theta\in(\frac{\pi}{2},\frac{3\pi}{2})$ respectively correspond to the scattering amplitudes measured be detectors placed at $x=+\infty$ and $x=-\infty$. Equations~(\ref{z82b}) and (\ref{z84b}), identify these with certain linear combinations of the scaled amplitudes, $\breve A_\pm$ and $\breve B_\pm$, and the delta function. This is the main reason for our use of these amplitudes.

As seen from (\ref{z81b}) -- (\ref{z84b}), the scattering problem for the potential has a solution, if there is a procedure to determine $\breve A_+$ and $\breve B_-$ for the $\breve A_-$ and $\breve B_+$ given by (\ref{z81b}). The scattering operator, more commonly known as the S-matrix, achieves this goal. We can represent it by a $2\times 2$ matrix $\widehat\bS$ that fulfills,
    \be
    \widehat\bS\left[\begin{array}{c}
    \breve A_-\\
    \breve B_+\end{array}\right]=
    \left[\begin{array}{c}
    \breve A_+\\
    \breve B_-\end{array}\right].
    \label{S-def-2D}
    \ee
This is the two-dimensional analog of the definition of the S-matrix in one dimension, namely (\ref{S-def-1D}). Notice, however, that unlike the latter, $\widehat\bS$ is a linear operator acting in the infinite-dimensional space of two-component functions,
    \[\sF_k^{2\times 1}:=\C^{2\times 1}\otimes\sF_k=\left\{\left[\begin{array}{c}\phi_+\\ \phi_-\end{array}\right]~\Bigg|~\phi_\pm\in\sF_k~\right\},\]
where $\C^{2\times1}$ stands the vector space of $2\times 1$ complex matrices.

Ref.~\cite{pra-2016} uses a direct extension of the definition of the transfer matrix in one dimension, namely (\ref{M-def-1D}), to arrive at a multidimensional notion of the transfer matrix. In two dimensions, this is done by identifying the transfer matrix with a $2\times 2$ matrix $\widehat\bM$ with operator entries $\widehat M_{ij}:\sF_k\to\sF_k$ such that
    \be
    \widehat\bM\left[\begin{array}{c}
    A_-\\
    B_-\end{array}\right]=
    \left[\begin{array}{c}
    A_+\\
    B_+\end{array}\right].
    \label{M-def}
    \ee
The utility of this transfer matrix in potential scattering relies on the following relations which in conjunction with (\ref{z82}) and (\ref{z84}) allow for the determination of the scattering amplitudes $f^{l/r}$.
    \begin{align}
    &A^l_+=2\pi \widehat M_{11}\:\delta_{p_0}+\widehat M_{12}B_-^l,
    \label{z201}\\
    &\widehat M_{22} B^l_-=-2\pi \widehat M_{21}\,\delta_{p_0},
    \label{z205}\\
    &A^r_+=\widehat M_{12}B^r_-,
    \label{z203}\\
    &\widehat M_{22} B^r_-=2\pi\,\delta_{p_0},
    \label{z206}
    \end{align}
where $\delta_{p_0}$ stands for the Dirac delta function centered at $p_0$, i.e., $\delta_{p_0}(p):=\delta(p-p_0)$, and $|p_0|<k$.
Equations (\ref{z201}) -- (\ref{z206}) follow from (\ref{M-def}) once we substitute (\ref{z81}) and (\ref{z83}) for $A_\pm$ and $B_\pm$, \cite{pra-2016,prsa-2016}. Notice that (\ref{z201}) and (\ref{z203}) reduce the determination of $A^{l/r}_+$ to that of $B^{l/r}_-$, while (\ref{z205}) and (\ref{z206}) provide linear integral equations for the latter.\footnote{In general, we can associate integral kernels $M_{ij}(p,q)$ to $\widehat M_{ij}$ and express (\ref{z205}) and (\ref{z206}) as the integral equations,
$\int_{-k}^k dq\: M_{22}(p,q) B^l_-(q)=-2\pi M_{21}(p,p_0)$ and $\int_{-k}^k dq\: M_{22}(p,q)B^r_-(q)=2\pi\,\delta(p-p_0)$, respectively. This follows from 
Eqs.~(\ref{M-expand-100}) and (\ref{bHb-100}) below, and the fact that $\widehat\sV(x)$ is the integral operator given by (\ref{v-def}).}

The above procedure for determining the scattering amplitudes using the entries of the transfer matrix admits a slightly simpler variation, if we employ the dimensionless coefficient functions $\breve A_\pm$ and $\breve B_\pm$ instead of $A_\pm$ and $B_\pm$, and make use of (\ref{z81b}) -- (\ref{z84b}). In light of (\ref{breves}), the role of $\widehat\bM$ is now played by
    \be
    \widehat{\breve\bM}:=\varpi(\hat p)\,\widehat{\bM}\,\varpi(\hat p)^{-1},
    \label{bMb-bM}
    \ee
where for every pair of functions $f,g:\R\to\C$, we define the function $f(\hat p)g$ by
    \[\big(f(\hat p)g\big)(p):=f(p)g(p).\]
It is easy to see that $\widehat{\breve\bM}$ is a transfer matrix (with operator entries $\widehat{\breve M}_{jl}$ acting in $\sF_k$) that satisfies
    \be
    \widehat{\breve\bM}
    \left[\begin{array}{c}
    \breve A_-\\
    \breve{B}_-\end{array}\right]=
    \left[\begin{array}{c}
    \breve A_+\\
    \breve B_+\end{array}\right].
    \label{M-def-b}
    \ee
With the help of this equation, we can express (\ref{z201}) -- (\ref{z206}) as
    \begin{align}
    &\breve A^l_+=2\pi\varpi(p_0)\,
    \widehat{\breve M}_{11}\:\delta_{p_0}+\widehat{\breve M}_{12}\breve B_-^l,
    \label{z201b}\\
    &\widehat{\breve M}_{22} \breve B^l_-=-2\pi\varpi(p_0)\,\widehat{\breve M}_{21}\,\delta_{p_0},
    \label{z205b}\\
    &\breve A^r_+=\widehat{\breve M}_{12}\breve B^r_-,
    \label{z203b}\\
    &\widehat{\breve M}_{22}\breve B^r_-=2\pi\varpi(p_0)\,\delta_{p_0}.
    \label{z206b}
    \end{align}
Again, (\ref{z205b}) and (\ref{z206b}) are linear integral equations for $\breve B^{l/r}_-$. Solving them and using the result in (\ref{z201b}) and (\ref{z203b}), we can determine $\breve A^{l/r}_+$. This yields a solution for the scattering problem by virtue of (\ref{z82b}) and (\ref{z84b}).

It is important to note that unlike its predecessors' \cite{pendry-1984,pendry-1990a, pendry-1990b,pendry-1994,mclean,ward-1996,pendry-1996}, the definition of the transfer matrix $\widehat{\breve\bM}$ does not involve a slicing or discretization of the configuration or momentum space variables. The same holds for the transfer matrix $\widehat{\bM}$. But $\widehat{\breve\bM}$ turns out to be more convenient to use.\footnote{This is because the entries of $\widehat{\breve\bM}$ enter the equations for scaled amplitudes, $\breve A_\pm$ and $\breve B_\pm$, and that the scattering amplitudes $\ff^{l/r}$ have simpler expressions in terms of these amplitudes than $A_\pm$ and $B_\pm$ (which are related by $\widehat{\bM}$.)} For this reason, we will refer to it as the ``fundamental transfer matrix.''

The introduction of the fundamental transfer matrix offers an alternative to the traditional methods of solving scattering problems. Its utility for this purpose involves the following steps.
    \begin{enumerate}
    \item Finding the fundamental transfer matrix $\widehat{\breve\bM}$.
    \item Solving (\ref{z205b}) and (\ref{z206b}) for $\breve B^{l/r}_-$.
    \item Substituting $\breve B^{l/r}_-$ in (\ref{z201b}) and (\ref{z203b}) to find $\breve A^{l/r}_+$.
    \item Substituting $\breve B^{l/r}_-$ and $\breve A^{l/r}_+$ in (\ref{z82b}) and (\ref{z84b}) to determine the scattering amplitudes $\ff^{l/r}$.
    \end{enumerate}
In Sec.~\ref{Sec3}, we derive a dynamical equation that yields a Dyson series expansion of $\widehat{\breve\bM}$. The solution of (\ref{z205b}) and (\ref{z206b}) requires the knowledge of the potential. These equations have a unique solution, if $\widehat{\breve M}_{22}$ has a nontrivial kernel (null space). The operator $\widehat{\breve M}_{22}$ is the two-dimensional counterpart of the $M_{22}$ entry of the transfer matrix $\bM$ in one dimension. Therefore, the condition that $\widehat{\breve M}_{22}$ has a nontrivial kernel is the two-dimensional counterpart of $M_{22}=0$. This equation is of particular interest, because the wavenumbers $k$ for which it holds correspond to the spectral singularities of the potential \cite{prl-2009}. This suggests a characterization of spectral singularities in two dimensions in terms of the nontriviality of  the kernel of $\widehat{\breve M}_{22}$.

It is also worth mentioning that (\ref{z201b}) -- (\ref{z206b}) reproduce their one-dimensional counterparts, namely (\ref{RRT-M}), if we replace $2\pi\varpi(p_0)\delta_{p_0}$ with 1 and identify $\breve A_+^l$, $\breve  B_+^l$, $\breve  A_+^r$, and $\breve  B_+^r$
respectively with the left transmission amplitude $T^l$, left reflection amplitude $R^l$, right reflection amplitude $R^r$, and right transmission amplitude $T^r$.

\section{Dynamical equation for fundamental transfer matrix}
\label{Sec3}

The formulation of the stationary scattering in terms of the fundamental transfer matrix $\widehat{\breve\bM}$ is useful provided that we can evaluate it and solve the integral equations (\ref{z205b}) and (\ref{z206b}) for $\breve B^{l/r}_-$. Because $\widehat{\breve\bM}$ and $\widehat{\bM}$ are related by a similarity transformation, the determination of any of these transfer matrices will yield the other. Ref.~\cite{pra-2016} shows that similarly to its one-dimensional analog \cite{ap-2014,pra-2014a} we can relate $\widehat\bM$ to the time-evolution operator for an effective non-unitary quantum system and derive a Dyson series expansion for it. In what follows we pursue a similar route that takes into account the role of the evanescent waves. This leads to a slightly different dynamical equation and Dyson series expansion for $\widehat\bM$ than those obtained in \cite{pra-2016}. As we show in the next section, an implicit assumption employed in \cite{pra-2016} corrects the discrepancy and the prescription proposed in \cite{pra-2016} for calculating the scattering amplitudes $\ff^{l/r}$ produces the correct result for the applications considered therein.

Let $\psi$ be the general bounded solution $\psi$ of the stationary Schr\"odinger equation (\ref{sch-eq}) that we consider in Sec.~\ref{Sec2}, and for each $x\in\R$, $\Psi_\pm(x):\R\to\C$ and $\bPsi(x):\R\to\C^{2\times 1}$ be the functions defined by
    \begin{align}
    &\big(\Psi_\pm(x))(p):=
    \frac{1}{2}\,e^{\pm i \varpi_r(p)x}\,
    \left[\varpi(p)\tilde\psi(x,p)\pm i\,\tilde\psi'(x,p)\right],
    &&\bPsi(x):=\left[\begin{array}{c}
    \Psi_-(x)\\
    \Psi_+(x)\end{array}\right],
    \end{align}
where
    \[\varpi_{\rm r}(p):=\RE[\varpi(p)]=\left\{\begin{array}{ccc}
    \sqrt{k^2-p^2} & \for & |p|< k\\
    0& \for & |p|\geq k
    \end{array}\right.=
    \left\{\begin{array}{ccc}
    \varpi(p) & \for & |p|< k,\\
    0& \for & |p|\geq k.
    \end{array}\right.\]
Then, in view of (\ref{sB=}) -- (\ref{asym-1}) and (\ref{breves}),
    \begin{align}
    &\big(\bPsi(x)\big)(p)=\left[\begin{array}{c}
    \breve A_-(p)\\
    \breve B_-(p)+\breve C_-(p)e^{|\varpi(p)|x}\end{array}\right]~~\for~~x\leq a_-,
    \label{asym-11}\\[6pt]
    &\big(\bPsi(x)\big)(p)=\left[\begin{array}{c}
    \breve A_+(p)+\breve C_+(p) e^{-|\varpi(p)|x}\\
    \breve B_+(p)\end{array}\right]~~\for~~x\geq a_+,
    \label{asym-12}
    \end{align}
where
    \be
    \breve C_\pm:=\varpi\,C_\pm.
    \label{breve-C}
    \ee
Equations~(\ref{asym-11}) and (\ref{asym-12}) imply
    \begin{align}
    &\lim_{x\to-\infty}\bPsi(x)=\left[\begin{array}{c}
    \breve A_-\\
    \breve B_-\end{array}\right],
    &&\lim_{x\to+\infty}\bPsi(x)=\left[\begin{array}{c}
    \breve A_+\\
    \breve B_+\end{array}\right].
    \label{asym-13}
    \end{align}
We can also use (\ref{ABC-bound}), (\ref{asym-11}), and (\ref{asym-12}) to show that, for $|p|<k$,
    \begin{align}
    &\big(\bPsi(x)\big)(p)=\left[\begin{array}{c}
    \breve A_-(p)\\
    \breve B_-(p)\end{array}\right]~~\for~~x\leq a_-,
    &&
    \big(\bPsi(x)\big)(p)=\left[\begin{array}{c}
    \breve A_+(p)\\
    \breve B_+(p)\end{array}\right]~~\for~~x\geq a_+.
    \label{asym-14}
    \end{align}

Next, we observe that (\ref{sch-eq-FT}) is equivalent to
    \be
    i\bPsi'(x)= \widehat{\breve\bH}(x)\bPsi(x),\quad\quad\quad x\in\R,
    \label{TD-sch-eq}
    \ee
where
    \begin{align}
    &\widehat{\breve\bH}(x):=\frac{1}{2}\:e^{-i\varpi_{\rm r}(\hat p)x\bsigma_3}
    \widehat\sV(x)\,\bcK
    \, e^{i\varpi_{\rm r}(\hat p)x\bsigma_3}\varpi(\hat p)^{-1}
    -i\varpi_{\rm i}(\hat p)\bsigma_3,
    \label{bH-def}\\
    &\bcK:=\left[\begin{array}{cc}
    1 & 1\\-1 & -1\end{array}\right],
    \end{align}
$\bsigma_3$ is the diagonal Pauli matrix, and
    \bea
    \varpi_{\rm i}(p)&:=&\IM[\varpi(p)]
    =\left\{\begin{array}{ccc}
    0 &\for&|p|<k\\
    \sqrt{p^2-k^2}&\for&|p|\geq k
    \end{array}\right.=\left\{\begin{array}{ccc}
    0 &\for&|p|<k,\\
    |\varpi(p)|&\for&|p|\geq k.
    \end{array}\right.\nn
    \eea
We may view (\ref{TD-sch-eq}) as the ``time-dependent'' Schr\"odinger equation for a non-stationary quantum system with the Hamiltonian operator (\ref{bH-def}) and $x$ playing the role of ``time.''

Let $\widehat{\breve\bU}(x,x_0)$ denote the evolution operator for the Hamiltonain (\ref{bH-def}) and an initial ``time'' $x_0$. This is the operator satisfying
    \begin{align}
    &i\partial_x\widehat{\breve\bU}(x,x_0)=\widehat{\breve\bH}(x)\widehat{\breve\bU}(x,x_0),
    &&\widehat{\breve\bU}(x_0,x_0)=\widehat\bI,
    \label{sch-eq-U}\\
    &\bPsi(x)=\widehat{\breve\bU}(x,x_0)\bPsi(x_0),
    \label{P-UP}
    \end{align}
where $\widehat\bI$ is the identity operator for the space $\sF^{2\times 1}:=\C^{2\times1}\otimes\sF$ of two-component state vectors, where $\sF$ is the space of complex-valued functions of $p$.\footnote{We can write $\widehat\bI$ as the product of the $2\times2$ identity matrix $\bI$ and the identity operator $\hat 1$ for $\sF$.}  In view of (\ref{M-def}), (\ref{asym-13}), and (\ref{P-UP}),
    \be
    \widehat{\breve\bM}=\lim_{x_\pm\to\pm\infty}\widehat{\breve\bU}(x_+,x_-).
    \label{M=U}
    \ee
We can express (\ref{sch-eq-U}) in the form of the following Dyson series \cite{sakurai}.
    \bea
    \widehat{\breve\bU}(x,x_0)&=&\widehat\bI+\sum_{n=1}^\infty (-i)^n
        \int_{x_0}^x \!\!dx_n\int_{x_0}^{x_n} \!\!dx_{n-1}
        \cdots\int_{x_0}^{x_2} \!\!dx_1\,
        \widehat{\breve\bH}(x_n)\widehat{\breve\bH}(x_{n-1})\cdots
        \widehat{\breve\bH}(x_1)\nn\\
        &=:&\sT\exp\left[-i\int_{x_0}^x dx'\:\widehat{\breve\bH}(x')\right],
        \label{U-expand}
    \eea
where $\sT$ denotes the time-ordering operation with $x$ playing the role of ``time.'' Substituting (\ref{U-expand}) in (\ref{M=U}) yields the Dyson series expansion of the fundamental transfer matrix;
    \bea
    \widehat{\breve\bM}&=&\widehat\bI+\sum_{n=1}^\infty (-i)^n
            \int_{-\infty}^\infty \!\!dx_n\int_{-\infty}^{x_n} \!\!dx_{n-1}
            \cdots\int_{-\infty}^{x_2} \!\!dx_1\,
            \widehat{\breve\bH}(x_n)\widehat{\breve\bH}(x_{n-1})\cdots\widehat{\breve\bH}(x_1)\nn\\
        &=:&\sT\exp\left[-i\int_{-\infty}^\infty dx\:\widehat{\breve\bH}(x)\right].
        \label{M-expand-b}
    \eea

In view of (\ref{bMb-bM}), we can use (\ref{M-expand-b}) to derive the following Dyson series expansion for the transfer matrix $\widehat{\bM}$.
    \bea
    \widehat{\bM}&=&\widehat\bI+\sum_{n=1}^\infty (-i)^n
            \int_{-\infty}^\infty \!\!dx_n\int_{-\infty}^{x_n} \!\!dx_{n-1}
            \cdots\int_{-\infty}^{x_2} \!\!dx_1\,
            \widehat{\bH}(x_n)\widehat{\bH}(x_{n-1})\cdots\widehat{\bH}(x_1),
        \label{M-expand-100}
    \eea
where
    \be
    \widehat{\bH}(x):=\varpi(\hat p)^{-1}\,\widehat{\breve\bH}(x)\,\varpi(\hat p)=
    \frac{1}{2}\:\varpi(\hat p)^{-1}e^{-i\varpi_{\rm r}(\hat p)x\bsigma_3}
    \widehat\sV(x)\,\bcK
    \, e^{i\varpi_{\rm r}(\hat p)x\bsigma_3}
    -i\varpi_{\rm i}(\hat p)\bsigma_3.
   \label{bHb-100}
    \ee

\section{Composition rule for fundamental transfer matrix}

In the absence of the potential, i.e., $v(x,y)=0$, we have $\widehat\sV(x)=\hat 0$, where $\hat 0$ is the zero operator acting in $\sF$, and $\widehat{\breve\bH}(x)=\widehat{\bH}_0:=-i\varpi_{\rm i}(\hat p)\bsigma_3$. We identify the latter with the ``free Hamiltonian,'' and examine the dynamics of the system in the interaction picture \cite{sakurai}, where the evolving state vectors are given by
    \be
    \bPhi(x):=e^{i\widehat{\bH}_0x}\bPsi(x)=e^{\varpi_{\rm i}(\hat p)x\bsigma_3}\bPsi(x).
    \label{bPhi=}
    \ee
These satisfy the ``time-dependent'' Schr\"odinger equation, $i\bPhi'(x)=\widehat{\breve\bcH}(x)\bPhi(x)$, for the ``interaction-picture Hamiltonian,''
    \bea
    \widehat{\breve\bcH}(x)&:=&e^{i\widehat{\bH}_0x}
    \widehat{\breve\bH}(x)e^{-i\widehat{\bH}_0x}-\widehat{\bH}_0\nn\\
    &=&\frac{1}{2}\,
    e^{-i\varpi (\hat p)x\bsigma_3}
    \widehat\sV(x)\,\bcK
    \, e^{i\varpi (\hat p)x\bsigma_3}\varpi(\hat p)^{-1}.
    \label{bcH-def}
    \eea
We can use the interaction-picture evolution operator $\widehat{\breve\bcU}(x,x_0)$, which is given by
    \be
    \widehat{\breve\bcU}(x,x_0):=e^{i\widehat{\bH}_0x}\,\widehat{\breve\bU}(x,x_0)
    e^{-i\widehat{\bH}_0x_0}=
    e^{\varpi_{\rm i}(\hat p)x\bsigma_3}\,\widehat{\breve\bU}(x,x_0)
    e^{-\varpi_{\rm i}(\hat p)x_0\bsigma_3},
    \label{bcU=1}
    \ee
to define the operator,
    \bea
    \widehat{\breve\bfM}&:=&\lim_{x_\pm\to\pm\infty}\widehat{\breve\bcU}(x_+,x_-)=
    \sT\exp\left[-i\int_{-\infty}^\infty dx\:\widehat{\breve\bcH}(x)\right].
    \label{bcMb-def1}
    \eea
We call this the ``auxiliary transfer matrix.'' It is related to the transfer matrix $\widehat{\bfM}$ constructed in Ref.~\cite{pra-2016} according to
    \bea
    \widehat{\bfM}&:=&\varpi(\hat p)^{-1}\widehat{\breve\bfM}\,\varpi(\hat p)=
    \sT\exp\left[-i\int_{-\infty}^\infty dx\:\widehat{\bcH}(x)\right],
    \label{bcM-def1}
    \eea
where
    \be
    \widehat{\bcH}(x):=\varpi(\hat p)^{-1}\widehat{\breve\bcH}\,\varpi(\hat p)
    =\frac{1}{2}\,\varpi(\hat p)^{-1}
    e^{-i\varpi (\hat p)x\bsigma_3}
    \widehat\sV(x)\,\bcK
    \, e^{i\varpi (\hat p)x\bsigma_3}.
    \ee
Note however that $\widehat{\bfM}\neq\widehat\bM$.

Next, consider slicing the space along a finite set of lines that are parallel to the $y$-axis. Let $\ell$ be a positive integer,  and $a_0,a_1,a_2,\cdots,a_\ell$ and $x_\pm$ are arbitrary real numbers such that
    \be
    x_-\leq a_-=a_0<a_1<a_2<\cdots<a_{\ell-1}<a_\ell=a_+\leq x_+.
    \label{as}
    \ee
Then, the interaction-picture evolution operator has the following semi-group property.
    \be
    \widehat{\breve\bcU}(x_+,x_-)=
    \widehat{\breve\bcU}(x_+,a_{\ell})\;
    \widehat{\breve\bcU}(a_\ell,a_{\ell-1})\;
    \widehat{\breve\bcU}(a_{\ell-1},a_{\ell-2})\;\cdots\;\widehat{\breve\bcU}(a_1,a_0)\;\widehat{\breve\bcU}(a_0,x_-).
    \label{semigroup-bcU-old}
    \ee
If $v(x,y)$ vanishes for a range of values of $x$, $\widehat{\sV}(x)=\hat 0$ and $\widehat{\breve\bcH}(x)=\widehat\bzero$, where $\widehat\bzero$ is the zero operator acting in $\sF^{2\times 1}$. This feature of $\widehat{\breve\bcH}(x)$ together with (\ref{semigroup-bcU-old}) are responsible for the composition property of the auxiliary transfer matrix \cite{pra-2016}. To see this, we let $v_j:\R\to\C$ be truncations of the potential $v$ given by
    \begin{align}
    &v_1(x,y):=\left\{\begin{array}{ccc}
    v(x,y) & \for & x\in[a_0, a_1],\\
    0 & \for & x\notin[a_0, a_1],\end{array}\right.
    &&v_{m+1}(x,y):=\left\{\begin{array}{ccc}
    v(x,y) & \for & x\in (a_{m},a_{m+1}],\\
    0 & \for & x\notin (a_{m},a_{m+1}],
    \end{array}\right.
    \label{vs}
    \end{align}
with $m\in\{1,2,\cdots,\ell-1\}$, so that $v_1+v_2+\cdots+v_\ell=v$. Let $\widehat{\sV}_j(x)$, $\widehat{\breve\bH}_j(x)$, $\widehat{\breve\bU}_j(x,x_0)$, $\widehat{\breve\bcH}_j(x)$, $\widehat{\breve\bcU}_j(x,x_0)$, and $\widehat{\breve\bfM}_j$ be respectively the analogs of $\widehat\sV(x)$, $\widehat{\breve\bH}(x)$, $\widehat{\breve\bU}(x,x_0)$, $\widehat{\breve\bcH}(x)$, $\widehat{\breve\bcU}(x,x_0)$, and $\widehat{\breve\bfM}$ for the potentials $v_j$. Then,
    \bea
    \widehat{\breve\bcH}_j(x)&=&\left\{\begin{array}{ccc}
    \widehat{\breve\bcH}(x)&\for& x\in [a_{j-1},a_j],\\
    \widehat\bzero&\for&x\notin[a_{j-1},a_j],\end{array}\right.\\[6pt]
    \widehat{\breve\bcU}_{j}(x,x_0)&=&
    \left\{\begin{array}{ccc}
    \widehat\bI&\for&x_0\leq x\leq a_{j-1},\\
    \widehat{\breve\bcU}(x,x_0)&\for& a_{j-1}\leq x_0\leq x\leq a_{j},\\
    \widehat\bI&\for& a_j\leq x_0\leq x.\end{array}\right.
    \label{z330}
    \eea
This together with $\widehat\bfM_j:=\lim_{x_\pm\to\pm\infty}\widehat{\breve\bcU}_j(x_+,x_-)$ imply
    \be
    \widehat{\breve\bfM_j}=\widehat{\breve\bcU}(a_j,a_{j-1}).
    \label{bcMj=}
    \ee
Furthermore, because $\widehat{\breve\bcU}(x_+,a_\ell)=\widehat{\breve\bcU}(a_0,x_-)=\widehat\bI$, we can express (\ref{semigroup-bcU-old}) as
    \be
    \widehat{\breve\bcU}(x_+,x_-)=
    \widehat{\breve\bcU_{\ell}}(a_\ell,a_{\ell-1})\;
    \widehat{\breve\bcU_{\ell-1}}(a_{\ell-1},a_{\ell-2})\;\cdots\;\widehat{\breve\bcU_1}(a_1,a_0)
    ~~\for~~\pm x_\pm\geq \pm a_\pm.
    \label{semigroup-bcU}
    \ee
Eqs.~(\ref{bcMb-def1}), (\ref{as}), (\ref{bcMj=}), and (\ref{semigroup-bcU}) lead to
    \be
    \widehat{\breve\bfM}=\widehat{\breve\bfM}_\ell\widehat{\breve\bfM}_{\ell-1}\cdots\widehat{\breve\bfM}_1.
    \label{compose-bcM}
    \ee

Next, we use the analog of (\ref{bcU=1}) for $v_j$, namely
    \be
    \widehat{\breve\bcU}_j(x,x_0)=
    e^{\varpi_{\rm i}(\hat p)x\bsigma_3}\,\widehat{\breve\bU}_j(x,x_0)\,
    e^{-\varpi_{\rm i}(\hat p) x_0\bsigma_3},
    \nn
    \ee
and (\ref{M=U}), (\ref{bcU=1}), (\ref{bcMj=}), and (\ref{compose-bcM}) to establish the following composition property of the fundamental transfer matrix.  \bea
    \widehat{\breve\bM}&=&
    \lim_{x_\pm\to\pm\infty}e^{-\varpi_{\rm i}(\hat p)x_+\bsigma_3}\,
    \widehat{\breve\bfM}_\ell\; \widehat{\breve\bfM}_{\ell-1}\;\cdots\;
    \widehat{\breve\bfM}_1\, e^{\varpi_{\rm i}(\hat p)x_-\bsigma_3}.
    \label{decompose1}
    \eea
In view of (\ref{compose-bcM}) this equation implies
    \be
    \widehat{\breve\bM}=
    \lim_{x_\pm\to\pm\infty}e^{-\varpi_{\rm i}(\hat p)x_+\bsigma_3}\,
    \widehat{\breve\bfM}\, e^{\varpi_{\rm i}(\hat p)x_-\bsigma_3}.
    \label{bM-bcM}
    \ee
Notice also that we can use (\ref{M=U}) and the semi-group property of the evolution operator $\widehat{\breve\bU}(x,x_0)$ to express the fundamental transfer matrix in the form,
    \bea
    \widehat{\breve\bM}&=&\lim_{x_\pm\to\pm\infty}
    \widehat{\breve\bU}(x_+,a_\ell)\;
    \widehat{\breve\bU}_{\ell}(a_\ell,a_{\ell-1})\;
    \widehat{\breve\bU}_{\ell-1}(a_{\ell-1},a_{\ell-2})\;
    \cdots\;
    \widehat{\breve\bU}_1(a_1,a_0)\;\widehat{\breve\bU}(a_0,x_-),
    \nn\\
    &=&\lim_{x_\pm\to\pm\infty}
    e^{-\varpi_{\rm i}(\hat p)x_+\bsigma_3}
    \widehat{\breve\bU}_{\ell}(a_\ell,a_{\ell-1})\;
    \cdots\;
    \widehat{\breve\bU}_1(a_1,a_0)
    e^{\varpi_{\rm i}(\hat p)x_-\bsigma_3},
    \label{decompose2}
    \eea
where we have made use of $\widehat{\breve\bU}(x_+,a_\ell)=e^{-\varpi_{\rm i}(\hat p)(x_+-a_\ell)\bsigma_3}$ and $\widehat{\breve\bU}(a_0,x_-)=e^{\varpi_{\rm i}(\hat p)(x_--a_0)\bsigma_3}$. These follow from the fact that for $x<a_0$ and $x>a_\ell$, $\widehat\sV(x)=\hat 0$, and $\widehat{\breve\bH}(x)=\widehat\bH_0=-i\varpi_{\rm i}(\hat p)\bsigma_3$.

The composition rule~(\ref{decompose2}) reduces the problem of obtaining the fundamental transfer matrix for the potential $v$ to the calculation of $\widehat{\breve\bU}_j(x_+,x_-)$. Alternatively, one may compute $\widehat{\breve\bfM}_j$ and utilize (\ref{decompose1}).

\section{Solution of the scattering problem using auxiliary transfer matrix}

Because of the presence of the second term on the right-hand side of (\ref{bH-def}), the Dyson series expansion for the Hamiltonian $\widehat{\breve\bcH}(x)$ has a simpler structure than that for $\widehat{\breve\bH}(x)$. This in turn suggests that it should be easier to compute $\widehat{\breve\bfM}$ and use it to determine the fundamental transfer matrix $\widehat{\breve\bM}$ using (\ref{bM-bcM}). Alternatively, we may try to express (\ref{z201}) -- (\ref{z206}) in terms of the entries of $\widehat{\breve\bfM}$ and obtain a solution for the scattering problem by solving these equations. In the following we examine the latter approach.

First, consider an arbitrary function $\phi:\R\to\C$ such that $|\phi(p)|$ does not diverge as $p\to\pm\infty$ faster than a polynomial, i.e., there is some $n\in\Z^+$ such that $\lim_{p\to\pm\infty}|\phi(p)/(1+|p|^n)=0$.\footnote{This in particular allows for identifying $\phi$ and its Fourier transform with tempered distributions.} Let $\phi_{\rm os},\phi_{\rm ev}:\R\to\C$ be defined by
    \[\phi_{\rm os}(p):=\left\{\begin{array}{ccc}
    \phi(p) & \for & |p|<k,\\
    0& \for & |p|\geq k,\end{array}\right.
    \quad\quad\quad\quad\quad\phi_{\rm ev}(p)=\phi(p)-\phi_{\rm os}(p).\]
Then, by virtue of (\ref{varpi}),
    \begin{align}
    &\lim_{x\to\infty}e^{- \varpi_{\rm i}(\hat p) x}\phi=\phi_{\rm os},
    &&\lim_{x\to\infty}e^{- \varpi_{\rm i}(\hat p) x}\phi_{\rm ev}=0.
    \label{proj1}
    \end{align}
These equations identify the operator,
    \be
    \widehat\Pi_k:=\lim_{x\to\infty}e^{-\varpi_{\rm i}(\hat p)x}
    =\lim_{x\to-\infty}e^{\varpi_{\rm i}(\hat p)x},
    \label{proj}
    \ee
with the projection operator that acts in $\sF$ according to
    \be
    \widehat\Pi_k\phi=\phi_{\rm os}.
    \label{Pi-def}
    \ee
In view of (\ref{ABC-bound}), (\ref{proj1}), and (\ref{proj}), we have
    \begin{align}
    &\widehat\Pi_k A_\pm=A_\pm, &&\widehat\Pi_k B_\pm=B_\pm, &&
    \widehat\Pi_k C_\pm=0.
    \label{proj2}
    \end{align}
These in turn imply
    \begin{align}
    &\widehat\Pi_k \breve\sB_-=\breve B_-, && \widehat\Pi_k\breve\sA_+=\breve A_+,
    && \widehat\Pi_k\breve C_\pm=0,
    \label{proj2}
    \end{align}
where
    \begin{align}
    &\breve\sB_-:=\varpi(\hat p)\sB_-,
    &&\breve\sA_+:=\varpi(\hat p)\sA_+,
    \end{align}
and we have made use of (\ref{sB=}),  (\ref{sA=}), (\ref{breves}), and  (\ref{breve-C}).

Next, we use (\ref{asym-11}), (\ref{asym-12}), and (\ref{bPhi=}) to show that
    \begin{align}
    &\lim_{x\to-\infty}\bPhi(x)=\Phi(a_-)=
    \left[\begin{array}{c}
    \breve A_-\\
    \breve \sB_-\end{array}\right],
    &&\lim_{x\to+\infty}\bPhi(x)=\Phi(a_+)=
    \left[\begin{array}{c}
    \breve \sA_+\\
    \breve B_+\end{array}\right].
    \end{align}
In view of (\ref{bcMb-def1}), these imply
    \begin{align}
    \widehat{\breve\bfM}
    \left[\begin{array}{c}
    \breve A_-\\
    \breve \sB_-\end{array}\right]=\left[\begin{array}{c}
    \breve \sA_+\\
    \breve B_+\end{array}\right].
    \label{bcM-TM}
    \end{align}
This relation together with (\ref{z82b}) and (\ref{z84b}) provide a route to the solution of the scattering problem. If we replace the role of $\breve B_-$, $\breve A_+$, and $\widehat{\breve\bM}$ in the derivation of (\ref{z201b}) -- (\ref{z206b}) with $\breve\sB_-$, $\breve\sA_+$, and $\widehat{\breve\bfM}$, and employ (\ref{bcM-TM}), we are led to
    \begin{align}
    &\breve\sA^l_+= 2\pi\varpi(p_0)\widehat{\breve\fM}_{11}\:\delta_{p_0}+\widehat{\breve\fM}_{12}\breve\sB_-^l,
    \label{cz201}\\
    &\widehat{\breve\fM}_{22}\,\breve\sB_-^l=-2\pi\varpi(p_0)\widehat{\breve\fM}_{21}\:\delta_{p_0},
    \label{cz202}\\
    &\breve\sA^r_+=\widehat{\breve\fM}_{12}\sB^r_-,
    \label{cz203}\\
    &\widehat{\breve\fM_{22}}\,\breve\sB^r_-=2\pi\varpi(p_0)\:\delta_{p_0}.
    \label{cz204}
    \end{align}

Determination of $\breve\sB_-^{l/r}$ requires the solution of (\ref{cz202}) and (\ref{cz204}). Solving these equations and substituting the result in (\ref{cz201}) and (\ref{cz203}) yield $\breve\sA_+^{l/r}$. We can obtain the coefficient functions $\breve A_+^{l/r}$ and $\breve B_-^{l/r}$ entering the formulas
(\ref{z82}) and (\ref{z84}) for the scattering amplitudes $\ff^{l/r}$ by affecting the projection operator $\widehat\Pi_k$ on $\breve\sB_-^{l/r}$ and $\breve\sA_+^{l/r}$, respectively;
    \begin{align}
    &\breve B_-^{l/r}=\widehat\Pi_k\breve\sB_-^{l/r},
    &&\breve A_+^{l/r}=\widehat\Pi_k\breve\sA_+^{l/r}.
    \end{align}
In practice, this means that we can obtain $\breve B_-^{l/r}(p)$ and $\breve A_+^{l/r}(p)$ by evaluating $\breve\sB_-^{l/r}(p)$ and $\breve\sA_+^{l/r}(p)$ at the values of $p$ such that $|p|<k$. This is simply because
    \begin{align}
    &\breve B_-^{l/r}(p)=\left\{\begin{array}{ccc}
    \breve\sB_-^{l/r}(p) & \for & |p|<k,\\
    0 &\for &|p|\geq k,\end{array}\right.
    &&\breve A_+^{l/r}(p)=\left\{\begin{array}{ccc}
    \sA_+^{l/r}(p) & \for & |p|<k,\\
    0 &\for &|p|\geq k.\end{array}\right.
    \end{align}

Next, we pursue an alternative approach for using $\widehat{\breve\bfM}$ as a tool for solving the scattering problem for the potential $v$. With the help of (\ref{M-def}) , (\ref{bM-bcM}), (\ref{proj}), (\ref{proj2}), and (\ref{bcM-TM}), we can show that
    \bea
    \bzero&=&\left[\begin{array}{c}
    \breve A_+\\
    \breve B_+\end{array}\right]-\widehat{\breve \bM}
    \left[\begin{array}{c}
    \breve A_-\\
    \breve B_-\end{array}\right]\nn\\
    &=&
    \left[\begin{array}{c}
    \breve A_+\\
    \breve B_+\end{array}\right]-
    \lim_{x\to\infty}e^{-\varpi_{\rm i}(\hat p)x\bsigma_3}
    \widehat{\breve\bfM}\left[\begin{array}{c}
    \breve A_-\\
    \breve B_-\end{array}\right]\nn\\
    &=&\left[\begin{array}{c}
    \breve A_+\\
    \breve B_+\end{array}\right]-\lim_{x\to\infty}e^{-\varpi_{\rm i}(\hat p)x\bsigma_3}
    \left(
    \widehat{\breve\bfM}\left[\begin{array}{c}
    \breve A_-\\
    \breve\sB_-\end{array}\right]-
    \widehat{\breve\bfM}\left[\begin{array}{c}
    0\\
    \breve C_-\end{array}\right]\right)\nn\\
    &=&\left[\begin{array}{c}
    \breve A_+\\
    \breve B_+\end{array}\right]-
    \lim_{x\to\infty}e^{-\varpi_{\rm i}(\hat p)x\bsigma_3}
    \left(
    \left[\begin{array}{c}
    \breve \sA_+\\
    \breve B_+\end{array}\right]-
    \widehat{\breve\bfM}\left[\begin{array}{c}
    0\\
    \breve C_-\end{array}\right]\right)\nn\\
    &=&\left[\begin{array}{c}
    \breve A_+\\
    \breve B_+\end{array}\right]-
    \left[\begin{array}{c}
    \widehat\Pi_k\breve \sA_+\\
    \breve B_+\end{array}\right]+
    \lim_{x\to\infty}e^{-\varpi_{\rm i}(\hat p)x\bsigma_3}
    \widehat{\breve\bfM}\left[\begin{array}{c}
    0\\
    \breve C_-\end{array}\right]\nn\\
    &=&
    \lim_{x\to\infty}e^{-\varpi_{\rm i}(\hat p)x\bsigma_3}
    \widehat{\breve \bfM}\left[\begin{array}{c}
    0\\
    \breve C_-\end{array}\right].\nn
    \eea
According to the latter equation,
    \begin{align}
    &\lim_{x\to\infty} e^{-\varpi_{\rm i}(\hat p)x}\widehat{\breve\fM}_{12}
    \breve C_-=0,
    &&\lim_{x\to\infty} e^{\varpi_{\rm i}(\hat p)x}\widehat{\breve\fM}_{22}
    \breve C_-=0,
    \end{align}
which mean that
    \begin{align}
    &\widehat\Pi_k\widehat{\breve\fM}_{12}\breve C_-=0,
    &&\widehat{\breve\fM}_{22}\breve C_-=0.
    \label{proj4}
    \end{align}
In view of the latter relation, we can write (\ref{bcM-TM}) as
    \be
    \widehat{\breve\bfM}
    \left[\begin{array}{c}
    \breve A_-\\
    \breve B_-\end{array}\right]=
    \left[\begin{array}{c}
    \breve \sA_+-\widehat{\breve\fM}_{12}\breve C_-\\
    \breve B_+\end{array}\right].
    \label{z721}
    \ee
If we apply,
    \be
    \widehat\bPi_k:=\widehat\Pi_k\bI,
    \label{bPi-def}
    \ee
to both sides of this equation and use (\ref{proj2}) and (\ref{proj4}), we can show that
    \be
    \widehat\bPi_k\widehat{\breve\bfM}\,\widehat\bPi_k
    \left[\begin{array}{c}
    \breve A_-\\
    \breve B_-\end{array}\right]=
    \widehat\bPi_k\widehat{\breve\bfM}
    \left[\begin{array}{c}
    \breve A_-\\
    \breve B_-\end{array}\right]=
    \left[\begin{array}{c}
    \breve A_+\\
    \breve B_+\end{array}\right].
    \label{M=M}
    \ee
Equations (\ref{M-def}) and (\ref{M=M}) suggest
    \be
    \widehat{\breve\bM}=\widehat\bPi_k\,\widehat{\breve\bfM}\,\widehat\bPi_k,
    \label{M=M2b}
    \ee
provided that we identify $\widehat{\breve\bM}$  as an operator acting in $\sF_k^{2\times 1}$. With the help of this identification and (\ref{z721}), we can also check the consistency of (\ref{bM-bcM}) and (\ref{M=M2b}). Furthermore, because $\widehat\Pi_k$ and $\varpi(\hat p)$ commute, we can use (\ref{bMb-bM}), (\ref{bcM-def1}), and (\ref{M=M2b}) to infer
    \be
    \widehat{\bM}=\widehat\bPi_k\,\widehat{\bfM}\,\widehat\bPi_k.
    \label{M=M2}
    \ee

Ref.~\cite{pra-2016} advocates the utility of the transfer matrix $\widehat\bM$ as a tool for solving scattering problems but fails to recognize its difference with $\widehat\bfM$. For this reason, it proposes to solve the scattering problem by employing (\ref{cz201}) --(\ref{cz204}) with $\sA_+^{l/r}$ and $\sB_-^{l/r}$ changed to $A_+^{l/r}$ and $B_-^{l/r}$. This discrepancy has escaped various analytical and numerical tests performed on the specific applications considered in Refs.~\cite{pra-2016,prsa-2016,ol-2017,pra-2019,jpa-2018}, because the final result of the calculation of the scattering amplitudes turns out to be correct. A careful analysis shows that for all of these applications the expression for $\widehat\bM$ can be recovered from that of $\widehat\bfM$ by replacing $\widehat\sV(x)$ with the operator $\widehat\Pi_k\widehat\sV(x)\widehat\Pi_k$. In view of (\ref{M=M2}), it is not difficult to see that letting $\widehat\Pi_k\widehat\sV(x)\widehat\Pi_k$ play the role of $\widehat\sV(x)$ in the Dyson series for $\widehat{\bfM}$ will produce $\widehat{\bM}$, if this series terminates after its second term, i.e.,
    \be
    \widehat{\bfM}=\widehat\bI-i\int_{-\infty}^\infty dx\: \widehat{\bcH}(x).
    \label{truncated-cM}
    \ee
As we show in Sec.~6, this happens for the delta-function potential in two dimensions.
In general, (\ref{truncated-cM}) does not hold, and one cannot obtain $\widehat{\bM}$ from the
expression for $\widehat{\bfM}$ by replacing $\widehat\sV(x)$ with $\widehat\Pi_k\widehat\sV(x)\widehat\Pi_k$.

For $v=0$, $\widehat{\breve\bfM}=\bI$, and (\ref{proj4}) implies $C_-=0$. At first sight this seems unjustified, because there is no reason why one should not be able to construct a solution $\psi$ of the free Schr\"odinger equation~(\ref{sch-eq}) such that $\psi(x,y)=C_-(p) e^{k\varpi(k)x+ipy}$ for $x<a_-$, where $p$ is a real number such that $|p|>k$, and $C_-$ is a nonzero function. Note, however, that because $v=0$, this solution satisfies $\psi(x,y)=C_-(k) e^{k\varpi(k)x+ipy}$ for all $x\in\R$. In particular, it diverges exponentially as $x\to+\infty$. The fact that (\ref{proj4}) reduces to $C_-=0$ for $v=0$ is a direct consequence of the requirement that $\psi$ is a bounded solution of (\ref{sch-eq}).


The application of the auxiliary and fundamental transfer matrices in solving the scattering problem for short-range potentials satisfying (\ref{supp-x}) does not seem to encounter any serious problems when we deal with short-range potentials violating (\ref{supp-x}) or potentials $v$ with an infinite range such that for each $y\in\R$, $v_y(x):=v(x,y)$ is a short-range potential in one dimension. A typical example is an infinite-range potential of the form,
    \[ v(x,y)=\left\{\begin{array}{ccc}
    \fz\, e^{i\alpha y}&\for & x\in[a_-,a_+],\\
    0 &\for & x\notin[a_-,a_+],\end{array}\right.\]
where $\alpha$ is a positive real parameter, and $\fz$ is a real or complex coupling constant \cite{pra-2017,berry-1998}. This is an exactly solvable potential that fails to satisfy (\ref{truncated-cM}). A careful examination of the solution of its scattering problem shows that one cannot use $\widehat\Pi_k\widehat\sV(x)\widehat\Pi_k$ in place of $\widehat\sV(x)$ to compute its scattering amplitude.

\section{Implicit regularization of delta-function potential in 2D}

Consider the potentials of the form,
    \begin{align}
    &v(x,y)=\delta(x)\fg(y),
    \label{delta}
    \end{align}
where $\fg:\R\to\C$ is a function with Fourier transform $\tilde\fg$, \cite{jpa-2018}. Then (\ref{v-def}) and (\ref{bcH-def}) give
    \begin{align}
    &\widehat\sV(x)=\delta(x)\, \widehat{\sG},
    &&\widehat{\breve\bcH}(x)=
    \frac{1}{2}\,\delta(x)
    \widehat{\sG}\,\varpi(\hat p)^{-1} \,\bcK,
    \label{v-def-delta}
    \end{align}
where
    \be
    (\widehat{\sG}\phi)(p)=\frac{1}{2\pi}\int_{-\infty}^\infty\!\!dq~\tilde\fg(p-q)\phi(q),
    \label{sG}
    \ee
and $\phi\in\sF$.\footnote{$\widehat{\sG}\phi$ gives the convolution of $\tilde g$ and $\phi$.} Because $\bcK^2=\bzero$, (\ref{v-def-delta}) implies $\widehat{\breve\bcH}(x_2)\widehat{\breve\bcH}(x_1)=\widehat\bzero$, the Dyson series (\ref{bcM-def1}) for $\widehat{\breve\bfM}$ terminates, and we find
    \be
    \widehat{\breve\bfM}=\bI-\frac{i}{2}\,
    \widehat{\sG} \,\varpi(\hat p)^{-1} \bcK
    =\left[\begin{array}{cc}
    \hat 1-\mbox{\large$\frac{i}{2}$}\,\widehat\sG\, \varpi(\hat p)^{-1}& -\mbox{\large$\frac{i}{2}$}\,\widehat\sG\,\varpi(\hat p)^{-1} \\[6pt]
    \mbox{\large$\frac{i}{2}$}\,\widehat\sG\,\varpi(\hat p)^{-1} & \hat 1+\mbox{\large$\frac{i}{2}$}\,\widehat\sG\, \varpi(\hat p)^{-1}
    \end{array}\right].
    \label{bcM-delta}
    \ee
Substituting this equation in (\ref{M=M2b}), we have
    \be
    \widehat{\breve\bM}=\widehat\bPi_k-\frac{i}{2}\,\widehat\Pi_k
    \widehat{\sG}\,\varpi(\hat p)^{-1}\widehat\Pi_k \, \bcK
    =\widehat\bPi_k-\frac{i}{2}\,\widehat\Pi_k
    \widehat{\sG}\,\widehat\Pi_k\varpi(\hat p)^{-1} \, \bcK.
    \label{bM-delta}
    \ee
Because $\widehat{\breve\bM}$ acts in $\sF_k^{2\times 1}$, we can replace the first $\widehat\bPi_k$ on the right-hand side of this equation by $\widehat\bI$ and observe that we can obtain it from (\ref{bcM-delta}) by replacing $\widehat\sG$ with $\widehat\Pi_k\widehat{\sG}\,\widehat\Pi_k$. In view of
(\ref{v-def-delta}) and (\ref{bcM-delta}), this shows that we can obtain $\widehat{\breve\bM}$ by letting $\widehat\Pi_k\widehat\sV(x)\,\widehat\Pi_k$ play the role of $\widehat\sV(x)$ in the calculation of $\widehat{\breve\bfM}$. Similarly, we can determine
$\widehat{\bM}$ from the expression for $\widehat{\bfM}$. This is the procedure used in \cite{jpa-2018} to obtain $\widehat{\bM}$ for the class of potentials given by (\ref{delta}).

For a variety of different choices for the function $\fg$, it is possible to find analytic closed-form expressions for the solutions of (\ref{z205}) and (\ref{z206}). These together with  (\ref{z82}), (\ref{z84}), (\ref{z201}), and (\ref{z203}) lead to an exact solution of the scattering problem for these potentials. Among these are the multi-delta-function potentials,
    \be
    v_N(x,y)=\delta(x)\sum_{n=1}^N\fz_n\delta(y-a_n),
    \label{delta-multi}
    \ee
for arbitrary $N\in\Z^+$, $\fz_n\in\C$, and $a_n\in\R$, as well as the Dirac comb potential,
    \[v_{DC}(x,y)=\fz\,\delta(x)\!\!\!\sum_{n=-\infty}^\infty\delta(y-n a),\]
where $\fz\in\C$ and $a\in\R^+$, \cite{jpa-2018}.

The standard treatment of the scattering problem for the delta-function potentials (\ref{delta-multi}) leads to divergent terms, and there are well-known regularization and renormalization procedures to derive a physically sensible expression for the scattering problem for these potentials \cite{mead,manuel,Adhikari1,Adhikari2,Mitra,Nyeo,Camblong,ap-2019}. The application of the fundamental transfer matrix to these potentials do not involve any divergent terms and produces the expression obtained by the standard approach. In the remainder of this section we examine the implicit regularization property of the fundamental transfer matrix. For simplicity of presentation we confine our attention to the delta-function potentials,
    \be
    v(x,y)=\fz\,\delta(x)\delta(y-a),
    \label{delta-1}
    \ee
with $\fz\in\C$ and $a\in\R$, which corresponds to setting $\fg(y)=\fz\,\delta(y-a)$.

For this choice of $\fg$, we have $\tilde\fg(p):=\cF_{y,p}\{g(y)\}=\fz\,e^{-iap}$. Substituting this equation in (\ref{sG}) and making use of (\ref{inv-Fourier}), we have
    \be
    (\widehat\sG\phi)(p):=\fz\,e^{-iap}\cF^{-1}_{q,a}\{\phi(q)\}.
    \label{sG-delta1}
    \ee
In view of (\ref{bcM-delta}) and (\ref{sG-delta1}), the entries of the auxiliary transfer matrix
satisfy,
    \be
    (\widehat{\breve\fM}_{j l}\phi)(p)=\phi(p)\,\delta_{jl}+\frac{(-1)^j\,{i} \fz\,f_\phi(a)\, e^{-{i}ap}}{2},
    \label{z931}
    \ee
where $\delta_{jl}$ is the Kronecker delta symbol, and
    \be
    f_\phi(a):=\cF^{-1}_{q,a}\left\{\varpi(q)^{-1}\phi(q)\right\}=
    \frac{1}{2\pi}\int_{-\infty}^\infty dq\,\frac{e^{iaq}\phi(q)}{\varpi(q)}.
    \label{c-phi}
    \ee

Because the delta-function potential (\ref{delta-1}) is invariant under the reflection about the $y$-axis, its left and right scattering amplitudes coincide; $\ff^l=\ff^r$. This allows us to confine our attention to the study of its scattering problem for a right-incident wave. Having determined auxiliary transfer matrix $\widehat{\breve\bfM}$, we first try to use it to calculate the right scattering amplitude $\ff^r$. This requires the solution of (\ref{cz204}) which in view of (\ref{z931}) takes the following explicit form.
    \be
    \breve\sB_-^r(p)=2\pi\varpi(p_0)\delta(p-p_0)-\frac{i\fz}{2}\,f(a)\,e^{-iap},
    \label{z875}
    \ee
where
    \be
    f(a):=f_{\breve\sB_-^r}\!(a)=
    \frac{1}{2\pi}\int_{-\infty}^\infty dq\,\frac{e^{iaq}\breve\sB_-^r(q)}{\varpi(q)}.
    \label{z876}
    \ee
We can try to determine $f(a)$ by substituting (\ref{z875}) in (\ref{z876}). This gives
    \be
    f(a)=\frac{e^{iap_0}}{\displaystyle 1+\frac{i\fz}{4\pi}\cF_{q,0}\{\varpi(q)^{-1}\}}.
    \label{z877}
    \ee
But
    \be
    \cF_{q,0}\{\varpi(q)^{-1}\}=\int_{-\infty}^\infty \frac{dq}{\sqrt{k^2-q^2}}=\infty.
    \label{z878}
    \ee
This is the same logarithmic singularity that arises in the standard treatment of the delta-function potentials in two dimensions \cite{mead,manuel,Adhikari1,Adhikari2,Mitra,Nyeo,Camblong,ap-2019}. We can remove it by a regularization of the integral in (\ref{z878}) followed by a renormalization of the coupling constant $\fz$.

The above calculation shows that at least for the delta-function potential (\ref{delta-1}) the application of the auxiliary transfer matrix faces the same complications as the standard methods based on the Lippmann-Schwinger equation.

Next, we examine the application of the fundamental transfer matrix (\ref{bM-delta}) in addressing the scattering problem for the delta-function potential (\ref{delta-1}). To do this, first we use (\ref{Pi-def}) and (\ref{sG-delta1}) to show that
    \be
    (\widehat\Pi_k\widehat\sG\,\widehat\Pi_k\phi)(p):=
    \frac{\chi_k(p)e^{-iap}}{2\pi}\int_{-k}^k dq\:e^{iaq}\phi(q),
    \label{sG-delta2}
    \ee
where
    \be
    \chi_k(p):=
    \left\{\begin{array}{ccc}
    1&\for&|p|<k,\\
    0&\for&|p|\geq k.\end{array}\right.
    \label{chi}
    \ee
Employing (\ref{sG-delta2}) in (\ref{bM-delta}) and assuming that $|p|<k$, we obtain
    \be
    (\widehat{\breve M}_{j l}\,\phi)(p)=\chi_k(p)\left[\phi(p)\delta_{j l}
    +\frac{(-1)^{j}\,i\fz\,g_\phi\, e^{-iap}}{2}\right],
    \label{z9312f}
    \ee
where
    \be
    g_\phi:=\frac{1}{2\pi}
    \int_{-k}^k dq\:\frac{e^{iaq}\phi(q)}{\varpi(q)}.\nn
    \ee

Now, we are in a position to solve (\ref{z206b}) for $\breve B_-^r$. In view of (\ref{z9312f}), this equation reads
    \be
    \breve B_-^r(p)=\chi_k(p)\left[2\pi\varpi(p_0)\delta(p-p_0)-\frac{i\fz}{2}\,g_{\breve\sB_-^r}\,e^{-iap}\right],
    \label{z875b}
    \ee
where
    \be
    g_{\breve\sB_-^r}\!:=
    \frac{1}{2\pi}\int_{-k}^k dq\,\frac{e^{iaq}\breve B_-^l(q)}{\varpi(q)}.
    \label{z876b}
    \ee
Again, we substitute (\ref{z875b}) in (\ref{z876b}) to determine $g_{\breve\sB_-^l}$. Because $\int_{-k}^k dq/\sqrt{k^2-q^2}=\pi$, this gives
    \be
    g_{\breve\sB_-^r}=\frac{4\, e^{iap_0}}{4+i\fz}.
    \label{z878b}
    \ee
Inserting this equation in (\ref{z875b}) and making use of (\ref{z84b}), (\ref{z203b}), (\ref{z9312f}), and (\ref{z878b}), we find
    \begin{align}
    &\breve B^r_-(p)=\chi_k(p)\left[2\pi\varpi(p_0)\delta(p-p_0)-\frac{2i\fz\, e^{-ia(p-p_0)}}{4+i\fz}\right],\nn\\
    &\breve A^r_+(p)=-\frac{i\fz}{2}\, \chi_k(p)\,g(a) e^{-iap}=
    -\frac{2i\fz\, \chi_k(p)\,e^{-ia(p-p_0)}}{4+i\fz},\nn\\
    &\ff^r(\theta)=-\sqrt{\frac{2}{\pi}}\,\frac{\fz}{4+i\fz}\,e^{-iak(\sin\theta-\sin\theta_0)}.\nn
    \end{align}
This expression for the scattering amplitude agrees with the one obtained in Ref.~\cite{pra-2016,jpa-2018}. The standard coupling constant normalization scheme also produces the same result albeit with $\fz$ replaced with the renormalized coupling constant \cite{ap-2019}.

Comparing the application of the auxiliary and fundamental transfer matrices for the treatment of the delta-function potential (\ref{delta-1}), we can explain the source of the implicit regularization feature of the latter. The presence of the projection operator $\widehat\bPi_k$ in (\ref{M=M2b}) has the effect of imposing a cut-off on $p$, namely $|p|<k$. Therefore the application of the fundamental transfer matrix involves an implicit cut-off regularization of the delta-function potentials (\ref{delta-1}). The same holds for the multi-delta-function potentials (\ref{delta-multi}).

\section{Perfect broadband invisibility in 2D}

Refs.~\cite{prsa-2016,ol-2017,pra-2019} use the transfer-matrix formulation of potential scattering proposed in Ref.~\cite{pra-2016} to construct complex potentials displaying perfect broadband omnidirectional and unidirectional invisibility. In this section, we discuss the application of the fundamental transfer matrix for the same purpose.

We have treated the scattering amplitudes $\ff$  and $\ff^{l/r}$ as functions of the scattering angle $\theta$. But they also depend on the incidence angle $\theta_0$ and the wavenumber $k$. Making the $\theta_0$- and $k$-dependence of the scattering amplitudes explicit, we can express (\ref{ff-f-def}) as
    \be
    \ff(\theta;\theta_0,k)=\left\{\begin{array}{ccc}
    \ff^l(\theta;\theta_0,k) &\for & -\frac{\pi}{2}<\theta_0<\frac{\pi}{2},\\[6pt]
    \ff^r(\theta;\theta_0,k) &\for & \frac{\pi}{2}<\theta_0<\frac{3\pi}{2}.
    \end{array}\right.
    \ee
If $\ff(\theta;\theta_0,k)=0$ for all $\theta_0\in[0,2\pi)$, we say that the potential has ``omnidirectional invisibility'' for wavenumber $k$. Similarly, we speak of  ``unidirectional invisibility'' for a wavenumber $k$ when one and only one of the following conditions holds \cite{prsa-2016}.
    \begin{align}
    &\ff^l(\theta;\theta_0,k)=0~~\mbox{for all~~$\theta_0\in(-\frac{\pi}{2},\frac{\pi}{2})$},\nn\\
    &\ff^r(\theta;\theta_0,k)=0~~\mbox{for all~~$\theta_0\in(\frac{\pi}{2},\frac{3\pi}{2})$}. \nn
    \end{align}
We refer to these cases as ``left'' and ``right unidirectional invisibility'' at $k$, respectively. We qualify these invisibility properties as being ``broadband,'' if they hold for a continuous range of values of $k$. We call them ``perfect,'' if we can realize them without invoking any approximation scheme. For example, if there is some $\alpha_\pm\in\R^+$ such that $\alpha_-<\alpha_+$, for all $k\in[\alpha_-,\alpha_+]$, $\ff^r(\theta;\theta_0,k)=0$ for all $\theta_0\in(\frac{\pi}{2},\frac{3\pi}{2})$ and  $\ff^l(\theta;\theta_0,k)\neq 0$ for some $\theta_0\in(\frac{\pi}{2},\frac{3\pi}{2})$, and we can verify these conditions without relying on an approximation scheme, we say that $v$ has perfect broadband unidirectional invisibility from the right \cite{prsa-2016,pra-2019}.

The fundamental transfer matrix provides a convenient characterization of invisible potentials. For example, according to (\ref{z82b}), (\ref{z84b}), and (\ref{z201b}) -- (\ref{z206b}), omnidirectional invisibility for a wavenumber $k$ corresponds to the requirement that,
    \be
    \widehat{\breve\bM}=\widehat\bI,
    \label{condi-1}
    \ee
for this particular value of $k$.\footnote{$\widehat{\breve\bM}$ and $\widehat{\breve\bfM}$ depend on $k$, but not on $\theta_0$.} Here and in what follows, we view $\widehat{\breve\bM}$ as an operator acting in $\sF_k^{2\times 1}$, and interpret $\bI$ as the identity operator for $\sF_k^{2\times 1}$.

In view of (\ref{bcMb-def1}) and (\ref{M=M2b}), (\ref{condi-1}) holds, if for all $n\in\Z^+$ and $x_1,x_2,\cdots,x_n\in\R$,
    \be
    \widehat\bPi_k\widehat{\breve\bcH}(x_n)
    \widehat{\breve\bcH}(x_{n-1})\cdots
    \widehat{\breve\bcH}(x_1)\widehat\bPi_k=\widehat\bzero.
    \label{condi-n}
    \ee
According to (\ref{bcH-def}) the entries of $\widehat{\breve\bcH}(x)$ are given by
    \[\widehat{\breve\cH}_{jl}(x)=\frac{(-1)^{j+1}}{2}\:e^{i(-1)^j x\varpi(\hat p)}\widehat\sV(x)\,e^{i(-1)^{l+1} x\varpi(\hat p)}
    \omega(\hat p)^{-1}.\]
This together with  (\ref{bPi-def}) and the fact that functions of $\hat p$ commute with the projection operator $\widehat\Pi_k$ show that (\ref{condi-n}) holds, if for arbitrary complex-valued functions $f_1,f_2,\cdots, f_{n-1}$ of $p$,
    \be
    \widehat\Pi_k \widehat{\sV}(x_n)\,f_{n-1}(\hat p)
    \widehat{\sV}(x_{n-1})f_{n-2}(\hat p)
    \widehat{\sV}(x_{n-2})\cdots\cdots f_{1}(\hat p)
    \widehat{\sV}(x_1)\widehat\Pi_k=\widehat 0.
    \label{condi-n2}
    \ee
For $n=1$, this is equivalent to
    \be
    \widehat\Pi_k\,\widehat\sV(x)\,\widehat\Pi_k=\hat 0.
    \label{condi-2}
    \ee
We can use (\ref{v-def}) and (\ref{proj}) to express (\ref{condi-2}) as
    \be
    \int_{-k}^k dq\:\tilde v(x,p-q)\phi(q)=0~~\for~~|p|<k~~{\rm and}~~\phi\in\sF_k.
    \label{condi-3}
    \ee
In light of the identity,
    \be
    \int_{-k}^k dq\:\tilde v(x,p-q)\phi(q)=\int_{p-k}^{p+k}
    d\fK\:\tilde v(x,\fK)\phi(p-\fK),
    \label{thm1-z1}
    \ee
and the fact that $|p|<k$ and $p-k\leq \fK\leq p+k$ imply $|\fK|<2k$, we can satisfy (\ref{condi-3}) by demanding that $\tilde v(x,\fK)=0$ for $|\fK|< 2k$. In particular, if there is some $\alpha\in\R^+$ such that
    \be
    \tilde v(x,\fK)=0~~\for~~\fK\leq 2\alpha,
    \label{condi-4}
    \ee
(\ref{condi-3}) and consequently (\ref{condi-2}) hold for all wavenumbers $k$ in the range $(0,\alpha]$, \cite{ol-2017}.

A less obvious consequence of (\ref{condi-4}) is that it implies (\ref{condi-n2}). This follows from a more general result.  As we show in the Appendix~A, if there are positive real numbers $\alpha$ and $\beta$ such that $k\leq\alpha$ and
    \be
    \tilde v(x,\fK)=0~~\for~~\fK\leq \beta,
    \label{condi-4beta-n}
    \ee
then
    \be
    \widehat{\sV}(x_n)f_{n-1}(\hat p)
    \widehat{\sV}(x_{n-1})f_{n-2}(\hat p)\widehat{\sV}(x_{n-2})\cdots f_{1}(\hat p)
    \widehat{\sV}(x_1)\widehat\Pi_k=0~~\for~~n\geq\frac{k+\alpha}{\beta}.
    \label{thm1-114n}
    \ee
In particular, because $\alpha\geq k$, (\ref{condi-n2}) and consequently (\ref{condi-n}) hold for $n\geq 2\alpha/\beta$. This together with (\ref{bcMb-def1}) and (\ref{M=M2b}) prove the following theorem.\\[6pt]
{\bf Theorem~1}: Let $\alpha,\beta\in\R^+$, and $v$ be a short-range potential such that $\tilde v(x,\fK)=0$ for $\fK\leq\beta$. Then, the following assertions hold for $k\in(0,\alpha]$.
    \begin{itemize}
    \item[a)] For $\beta\geq 2\alpha$, $\widehat{\breve\bM}=\widehat\bI$.
    \item[b)] For $0<\beta<2\alpha$,
    \be
    \widehat{\breve\bM}=\widehat\bI+\sum_{n=1}^{\lceil 2\alpha/\beta-1\rceil} (-i)^n\!\!
            \int_{x_0}^x \!\!dx_n\int_{x_0}^{x_n} \!\!dx_{n-1}
            \cdots\int_{x_0}^{x_2} \!\!dx_1\,\widehat\bPi_k
            \widehat{\breve\bcH}(x_n)\widehat{\breve\bcH}(x_{n-1})\cdots
            \widehat{\breve\bcH}(x_1)\widehat\bPi_k,
            \label{Thm-01}
        \ee
    \end{itemize}
where $\widehat\bI$ is to be interpreted as the identity operator acting in $\sF_k^{2\times 1}$, and $\lceil x\rceil$ stands for the smallest integer not smaller than $x$.\vspace{6pt}

For a potential satisfying (\ref{condi-4}), $\beta=2\alpha$, and
Theorem~1 implies that $\widehat{\breve\bM}=\widehat\bI$ for $k\in(0,\alpha]$. This proves the following theorem on perfect broadband omnidirectional invisibility \cite{ol-2017}. \\[6pt]
{\bf Theorem~2}: Let $\alpha$ be a positive real number, and $v$ be a short-range potential such that $\tilde v(x,\fK)=0$ for $\fK\leq 2\alpha$. Then $v$ is omnidirectionally invisible for every wavenumber $k$ that does not exceed $\alpha$.\\[6pt]
Ref.~\cite{ol-2017} provides concrete examples of potentials satisfying the hypothesis of Theorem~2 and discusses their optical realizations in effectively two-dimensional isotropic media. This provides the first examples of complex permittivity profiles that display perfect broadband omnidirectional invisibility for transverse electric or transverse magnetic incident waves. Ref.~\cite{jpa-2020} reports a three-dimensional extension of this result that allows for the construction of isotropic media possessing perfect broadband omnidirectional invisibility for incident waves of arbitrary polarization.

Next, consider a potential satisfying (\ref{condi-4beta-n}) for $\beta=\alpha$, i.e.,
    \be
    \tilde v(x,\fK)=0~~\for~~\fK\leq \alpha,
    \label{condi-4alpha}
    \ee
and suppose that $k\in(0,\alpha]$. Then, (\ref{thm1-114n}) implies
    \be
    \widehat{\sV}(x_2)f_{1}(\hat p)
    \widehat{\sV}(x_{1})\widehat\Pi_k=\hat 0,
    \label{thm1-115}
    \ee
and Theorem~1 gives the following expression for the fundamental transfer matrix.
    \be
    \widehat{\breve\bM}=\widehat\bI-i\int_{-\infty}^\infty dx\:\widehat\bPi\,
    \widehat{\breve\bcH}(x)\,\widehat\bPi=
    \widehat\bI-\frac{i}{2}\int_{-\infty}^\infty dx\:
    e^{-i\varpi (\hat p)x\bsigma_3}
    \widehat\Pi\widehat\sV(x)\widehat\Pi\,\bcK
    \, e^{i\varpi (\hat p)x\bsigma_3}\varpi(\hat p)^{-1}.
    \label{thm3-1}
    \ee
This relation justifies the use of $\widehat\Pi\widehat\sV(x)\widehat\Pi$ instead of $\widehat\sV(x)$ in Ref.~\cite{pra-2019}.

In view of (\ref{thm1-115}) and (\ref{thm3-1}), for all $\phi\in\sF_k$, and $i_1,i_2,j_1,j_2\in\{1,2\}$,
    \be
    \big(\widehat{\breve M}_{i_1 j_1}-\delta_{i_1 j_1}\hat 1\big)
    \big(\widehat{\breve M}_{i_2 j_2}-\delta_{i_2 j_2}\hat 1\big)\phi=0.
    \label{thm3-2}
    \ee
As noted in \cite{pra-2019}, we can use (\ref{thm3-2}) to obtain a closed form expression for the scattering amplitudes $\ff^{l/r}$. For completeness we summarize the derivation of this expression. First, we write (\ref{z205b}) and (\ref{z206b}) as
    \bea
    \breve B^l_-&=&-\big(\widehat{\breve M}_{22}-\hat 1\big) \breve B^l_--2\pi\varpi(p_0)\,\widehat{\breve M}_{21}\,\delta_{p_0},
    \label{z205bb}\\
    \breve B^r_-&=&-\big(\widehat{\breve M}_{22}-\hat 1\big) \breve B^r_-+2\pi\varpi(p_0)\,\delta_{p_0}.
    \label{z206bb}
    \eea
We can then use (\ref{thm3-2}) and the fact that $\breve B^{l/r}_-$, $\delta_{p_0}$, and $\widehat{\breve M}_{21}\,\delta_{p_0}$ belong to $\sF_k$ to check that
    \begin{align}
    &\breve B^l_-=-2\pi\varpi(p_0)\,\widehat{\breve M}_{21}\,\delta_{p_0},
    &&\breve B^r_-=-2\pi\varpi(p_0)\big(\widehat{\breve M}_{22}-2\hat 1)\,\delta_{p_0},
    \label{thm3-4}
    \end{align}
solve (\ref{z205bb}) and (\ref{z206bb}). Substituting (\ref{thm3-4}) in (\ref{z201b}) and (\ref{z203b}) and employing (\ref{thm3-2}), we obtain
    \begin{align}
    &\breve A^l_+=2\pi\varpi(p_0)\,\widehat{\breve M}_{11}\,\delta_{p_0},
    &&\breve A^r_+=2\pi\varpi(p_0)\,\widehat{\breve M}_{12}\,\delta_{p_0}.
    \label{thm3-5}
    \end{align}
If we insert (\ref{thm3-4}) and  (\ref{thm3-5}) in (\ref{z82b}) and (\ref{z84b}) and use (\ref{thm3-1}) to evaluate the $\widehat{\breve M}_{jl}\,\delta_{p_0}$ that appear in the resulting expressions for the scattering amplitudes, we are led to the following most remarkable formula \cite{pra-2019}.
    \be
    \ff^{l/r}(\theta;\theta_0,k)=-\frac{\tilde{\tilde v}\big(k(\cos\theta-\cos\theta_0),
    k(\sin\theta-\sin\theta_0)\big)}{2\sqrt{2\pi}},
    \label{Born1}
    \ee
where $\tilde{\tilde v}(\fK_x,\fK_y)$ stands for the two-dimensional Fourier transform of $v(x,y)$, i.e.,
    \[\tilde{\tilde v}(\fK_x,\fK_y):=\int_{-\infty}^\infty\!\! dx\int_{-\infty}^\infty\!\! dy\:e^{-i(\fK_x x+\fK_y y)}v(x,y).\]
Because the right-hand side of (\ref{Born1}) is precisely the formula one obtains for the scattering amplitude in the first Born approximation, the above calculation proves the following theorem that was originally reported in \cite{pra-2019}.\\[6pt]
{\bf Theorem~3}: Let $\alpha$ be a positive real number, and $v$ be a short-range potential such that $\tilde v(x,\fK)=0$ for $\fK\leq \alpha$. Then the first Born approximation gives the exact expression for the scattering amplitude of $v$ for wavenumbers $k$ not exceeding $\alpha$.\\[6pt]
Ref.~\cite{pra-2019} provides concrete examples of potentials fulfilling the hypothesis of this theorem and characterizes particular classes of such potentials that display perfect broadband unidirectional invisibility.

\section{Generalization to three dimensions}

The developments we report in Secs.~2-5 admit a straightforward three-dimensional generalization, where the scattering phenomenon is defined through the stationary Schr\"odinger equation,
    \be
    [-\nabla^2+v(\bfr)]\psi(\bfr)=k^2\psi(\bfr),
    \label{sch-eq-3D}
    \ee
$\nabla^2$ stands for the three-dimensional Laplacian, $v:\R^3\to\C$ is a short-range potential, $\bfr=x\,\bfe_x+y\,\bfe_y+z\,\bfe_z$ is the position vector, $\bfe_j$ is the unit vector along the $j$-axis, and $j\in\{x,y,z\}$. Following the standard convention, we consider situations where the source of the incident wave is located on either of the planes $z=-\infty$ or $z=+\infty$. These respectively correspond to the scattering of left- and right-incident waves.

The standard formulation of stationary scattering \cite{yafaev} relies on the existence of the so-called scattering solutions of (\ref{sch-eq-3D}). By definition, these satisfy
    \[\psi(\bfr)= e^{i\bk_0\cdot\bfr}+\frac{\ff(\vartheta,\varphi)\, e^{ikr}}{r}+o(r^{-1})~~\for~~r\to\infty,\]
where $\bk_0$ is the incident wave vector, $(r,\vartheta,\varphi)$ are the spherical coordinates of $\bfr$, and $\ff(\vartheta,\varphi)$ is the scattering amplitude. We can quantify the direction of $\bk_0$ by the spherical angles $\vartheta_0$ and $\varphi_0$. Because $|\bk_0|=k$, $(k,\vartheta_0,\varphi_0)$ are the spherical coordinates of $\bk_0$. For left- and right-incident waves, $\vartheta_0$ takes values in the intervals $[0,\frac{\pi}{2})$ and $(\frac{\pi}{2},\pi]$, and we label the corresponding scattering amplitudes by $\ff^l$ and $\ff^r$, respectively. In other words,
    \be
    \ff(\vartheta,\varphi)=\left\{\begin{array}{ccc}
    \ff^l(\vartheta,\varphi)&\for&\vartheta_0\in[0,\frac{\pi}{2}),\\
    \ff^r(\vartheta,\varphi)&\for&\vartheta_0\in(\frac{\pi}{2},\pi].\end{array}\right.
    \label{ff=ff-3D}
    \ee

Given a vector $\bu\in\R^3$ with components $u_x, u_y$, and $ u_z$, let $\vec u$ denote the projection of $\bu$ onto the $x$-$y$ plane, i.e., $\vec u:= u_x\bfe_x+ u_y\bfe_y$. In the following, we identify $\vec u$ and $\bu$ respectively with $(u_x,u_y)$ and $(\vec u,u_z)$, e.g.,
    \begin{align*}
    &\vec r=(x,y), &&\bfr=(\vec r,z).
    \end{align*}

Suppose that for $z\to\pm\infty$, $|v(\vec r,z)|$ tends to zero with such a rate that every bounded solution of (\ref{sch-eq-3D}) satisfies
    \be
    \psi(\vec r,z)\to\int_{\sD_k}\frac{d^2\vec p}{4\pi^2\varpi(\vec p)}\: e^{i\vec p\cdot\vec r}
    \left[\breve A_\pm(\vec p)e^{i\varpi(\vec p)z}+\breve B_\pm(\vec p)e^{-i\varpi(\vec p)z}\right]~~\for~~z\to\pm\infty,
    \ee
where
    \begin{align}
    &\sD_k:=\left\{\vec p\in\R^2~|~|\vec p|<k~\right\},
    &&\varpi(\vec p):=\left\{\begin{array}{ccc}
    \sqrt{k^2-|\vec p|^2}&\for&|\vec p|<k,\\[3pt]
    i\sqrt{|\vec p|^2-k^2}&\for&|\vec p|\geq k,\end{array}\right.\nn
    \end{align}
and $\breve A_\pm$ and $\breve B_\pm$ are functions of $\vec p\in\R^2$ that vanish for $|\vec p|\geq k$. In analogy to two dimensions, we use $\sF$ to denote the set of functions of $\vec p$, and let $\sF_k:=\{\phi\in\sF~|~\phi(\vec p)=0~\for~|\vec p|\geq k~\}$. Then, $\breve A_\pm,\breve B_\pm\in\sF_k$, and we identify the fundamental transfer matrix $\widehat{\breve\bM}$ with the linear operator acting in $\sF_k^{2\times 1}:=\C^{2\times 1}\otimes\sF_k$ that relates $\breve A_\pm$ and $\breve B_\pm$ via (\ref{M-def-b}). This equation leads to the following analogs of (\ref{z201b}) -- (\ref{z206b}).
    \begin{align}
    &\breve A^l_+=4\pi^2\varpi(\vec p_0)\,
    \widehat{\breve M}_{11}\:\delta_{\vec p_0}+\widehat{\breve M}_{12}\breve B_-^l,
    \label{z201b-3d}\\
    &\widehat{\breve M}_{22} \breve B^l_-=-4\pi^2\varpi(\vec p_0)\,\widehat{\breve M}_{21}\,\delta_{\vec p_0},
    \label{z205b-3d}\\
    &\breve A^r_+=\widehat{\breve M}_{12}\breve B^r_-,
    \label{z203b-3d}\\
    &\widehat{\breve M}_{22}\breve B^r_-=4\pi^2\varpi(\vec p_0)\,\delta_{\vec p_0},
    \label{z206b-3d}
    \end{align}
where
    \be
    \vec p_0:=k\sin\vartheta_0(\cos\varphi_0\,\bfe_x+\sin\varphi_0\,\bfe_y),
    \label{p0=}
    \ee
and $\delta_{\vec p_0}$ stands for the delta function centered at $\vec p_0$ in two dimensions, i.e., $\delta_{\vec p_0}(\vec p):=\delta(\vec p-\vec p_0)$.

The link between the fundamental transfer matrix and the left/right scattering amplitudes is provided by Eqs.~(\ref{z201b-3d}) -- (\ref{z206b-3d}) and the following identity which follows from an argument given in \cite[Appendix F]{pra-2016}.
    \bea
    \ff^l(\vartheta,\varphi)&=&-\frac{i}{2\pi}\times\left\{
    \begin{array}{ccc}
    \breve A^l_+(\vec p)-4\pi^2\varpi(\vec p_0)\delta(\vec p-\vec p_0)&\for&\vartheta\in[0,\frac{\pi}{2}),\\
    \breve B^l_-(\vec p)&\for&\vartheta\in(\frac{\pi}{2},\pi],\end{array}\right.
    \label{FL-3d}\\
    \ff^r(\vartheta,\varphi)&=&-\frac{i}{2\pi}\times\left\{
    \begin{array}{ccc}
    \breve A^r_+(\vec p)&\for&\vartheta\in[0,\frac{\pi}{2}),\\
    \breve B^r_-(\vec p)-4\pi^2\varpi(\vec p_0)\delta(\vec p-\vec p_0)&\for&\vartheta\in(\frac{\pi}{2},\pi],\end{array}\right.
    \label{FR-3d}
    \eea
where $\vec p_0$ and $\vec p$ are respectively given by (\ref{p0=}) and
    \be
    \vec p:=k\sin\vartheta(\cos\varphi\,\bfe_x+\sin\varphi\,\bfe_y).
    \label{p=3d}
    \ee

In practice, it is easier to determine the auxiliary transfer matrix $\widehat{\breve\bfM}$ and obtain the fundamental transfer matrix $\widehat{\breve\bM}$ using $\widehat{\breve\bM}=\widehat\bPi_k\, \widehat{\breve\bfM}\:
\widehat\bPi_k$, where $\widehat\bPi_k:=\widehat\Pi_k\bI$ and $\widehat\Pi_k$ is the projection operator mapping $\sF$ onto $\sF_k$ via
    \[\big(\widehat\Pi_k f\big)(p):=\left\{\begin{array}{ccc}
    f(p)&\for&|\vec p|<k,\\
    0 &\for&|\vec p|\geq k.\end{array}\right.\]
We identify the auxiliary transfer matrix with the operator,
    \be
    \widehat{\breve\bfM}:=\lim_{z_\pm\to\pm\infty}\widehat{\breve\bcU}(z_+,z_-)=
    \sT\exp\left[-i\int_{-\infty}^\infty dz\:\widehat{\breve\bcH}(z)\right],
    \ee
where $\widehat{\breve\bcU}(z,z_0)$ is the evolution operator for a quantum system with Hamiltonian operator,
    \be
    \widehat{\breve\bcH}(z):=\frac{1}{2}\,
    e^{-i\varpi (\,\widehat{\vec p}\,)z\bsigma_3}
    \widehat\sV(z)\,\bcK
    \, e^{i\varpi (\,\widehat{\vec p}\,)z\bsigma_3}\varpi(\,\widehat{\vec p}\,)^{-1},
    \label{bcH-def-ed}
    \ee
$z$ plays the role of ``time,'' $z_0$ is an initial value of $z$, for all $f,g\in\sF$ we have $\big(f(\,\widehat{\vec p}\,)g\big)(\vec p):=f(\vec p)g(\vec p)$,
    \be
    \big(\widehat\sV(z) f\big)(\vec p):=\frac{1}{4\pi^2}\int_{\R^2}\! d^2\vec q\:\tilde{\tilde v}(\vec p-\vec q,z) f(\vec q),
    \label{eq-149}
    \ee
and $\tilde{\tilde v}(\vec p,z):=\int_{\R^2}\!  d^2\vec r\:e^{i\vec p\cdot\vec r}v(\vec r,z)$ is the two-dimensional Fourier transform of $v(\vec r,z)$ over $\vec r$.

The application of the fundamental transfer matrix for the solution of the scattering problem for the delta-function potential in three dimensions,
    \be
    v(\bfr)=\fz\,\delta(\bfr),
    \label{delta-3D}
    \ee
avoids the singularities arising in the standard treatment of this potential and produces the same result for the scattering amplitudes $\ff^{l/r}$ as the one obtained by a regularization of the singularities and the renormalization of the coupling constant in the standard treatment. To see this, we consider the following three-dimensional analog of (\ref{delta}).
    \be
    v(x,y,z)=\fz\,\fg(\vec r)\,\delta(z),
    \label{delta-3D-1}
    \ee
where $\fg:\R^2\to\C$ is a function with (two-dimensional) Fourier transform $\tilde{\tilde{\fg}}$.

Pursuing the approach of Sec.~6, we can verify that the fundamental transfer matrix for the potential (\ref{delta-3D-1}) is given by (\ref{bM-delta}), where
    \be
    (\widehat\Pi_k\widehat{\sG}\,\widehat\Pi_k\phi)(\vec p)=\frac{\chi_k(\vec p)}{4\pi^2}\int_{\sD_k}d^2\vec q~\tilde{\tilde{\fg}}(\vec p-\vec q)\phi(\vec q),
    \label{delta-3D-2}
    \ee
and
    \[\chi_k(\vec p):=\left\{\begin{array}{ccc}
    1 &\for&|\vec p|< k,\\
    0 &\for&|\vec p|\geq k.\end{array}\right.\]
For $\fg(\vec r)=\delta(\vec r)$, (\ref{delta-3D-1}) reduces to (\ref{delta-3D}), and (\ref{delta-3D-2}) becomes $(\widehat\Pi_k\widehat{\sG}\,\widehat\Pi_k\phi)(\vec p)=(4\pi^2)^{-1}\chi_k(\vec p)\int_{\sD_k}d^2\vec q~\phi(\vec q)$. Substituting this equation in (\ref{bM-delta}) we find the following expression for the entries of the fundamental transfer matrix.
    \be
    \big(\widehat{\breve M}_{jl}\phi\big)(\vec p)=\chi_k(\vec p)\left[\delta_{jl}\phi(\vec p)
    +\frac{(-1)^j\; {i}\fz\,h_\phi}{2}\right],
    \label{delta-3D-3}
    \ee
where
    \be
    h_\phi:=\frac{1}{4\pi^2}\int_{\sD_k} d^2\vec q\: \frac{\phi(\vec q)}{\varpi(\vec q)}.
    \label{delta-3D-4}
    \ee

Next, we use (\ref{delta-3D-3}) to express (\ref{z206b-3d}) as
    \be
    \breve B^r_-(\vec p)=4\pi^2\varpi(\vec p_0)\delta(\vec p-\vec p_0)-\frac{i\fz\, h_{\breve  B^r_-}}{2}.
    \label{delta-3D-5}
    \ee
We can compute $h_{\breve  B^r_-}$ by substituting this equation in (\ref{delta-3D-4}). The result is
    \be
    h_{\breve  B^r_-}=\frac{4\pi}{4\pi+i k\,\fz},
    \label{delta-3D-6}
    \ee
where we have used the identity $\int_{\sD_k}d^2\vec q/\sqrt{k^2-q^2}=2\pi k$. Inserting (\ref{delta-3D-6}) in (\ref{delta-3D-5}) and using (\ref{z203b-3d}), (\ref{FR-3d}), (\ref{delta-3D-3}), (\ref{delta-3D-6}), and $\ff^l=\ff^r$, which follows from the invariance of the potential (\ref{delta-3D}) under the transformation, $z\to-z$, we have
    \bea
    &&\breve B^r_-(\vec p)=\chi_k(\vec p)\left[4\pi^2\varpi(\vec p_0)\delta(\vec p-\vec p_0)-\frac{2\pi i\,\fz}{4\pi+ik\,\fz}\right],\nn\\
    &&\breve A^r_+(\vec p)=-\frac{2\pi i\,\fz\,\chi_k(\vec p)}{4\pi+ik\,\fz},\nn\\
    &&\ff^l(\vartheta,\varphi)=\ff^r(\vartheta,\varphi)=-\frac{\fz}{4\pi+ik\,\fz}.\nn
    \eea
The standard treatment of the delta-function potential~(\ref{delta-3D}) which involves a regularization of a polynomial singularity and the renormalization of the coupling constant gives the same expression for $\ff^{l/r}$ with $\fz$ changed to the renormalized coupling constant. For the cases where $\fz$ is real, there is an argument due to Rajeev which leads to the same result \cite{Rajeev-1999}. Again our treatment avoids singular terms, for the solution of the scattering problem using the fundamental transfer matrix has a built-in cut-off regularization property.

We conclude this section by pointing out that we can easily generalize the results of Sec.~7 to three dimensions and establish the following three-dimensional analogs of Theorems~2 and 3.\\[6pt]
{\bf Theorem~4}: Let $\alpha$ be a positive real number, $\vec e$ be a unit vector lying in the $x$-$y$ plane, and $v:\R^3\to\C$ be a short-range potential such that $\tilde{\tilde v}(\vec\fK,z)=0$ for $\vec\fK\cdot\vec e\leq 2\alpha$. Then $v$ is omnidirectionally invisible for every wavenumber $k$ that does not exceed $\alpha$.\\[6pt]
{\bf Theorem~5}: Let $\alpha$ be a positive real number, $\vec e$ be a unit vector lying in the $x$-$y$ plane, and $v$ be a short-range potential such that $\tilde{\tilde v}(\vec\fK,z)=0$ for $\vec\fK\cdot\vec e\leq \alpha$. Then the first Born approximation gives the exact expression for the scattering amplitude of $v$ for wavenumbers $k\leq\alpha$.\\[6pt]
We outline the proofs of these theorems in Appendix~B.

\section{Concluding remarks}

The idea of making use of an analog of the transfer matrix to describe the propagation and scattering of waves in dimensions higher than one dates back to the 1980's \cite{pendry-1984}. The transfer matrices used for this purpose are obtained by slicing the space along one direction, discretizing the spatial or momentum variable(s) along the normal directions, assigning a transfer matrix for each slice, and multiplying the transfer matrices for the slices according to an analog of the composition rule for the transfer matrix in one dimension \cite{pendry-1996}. The outcome is a large numerical transfer matrix whose manipulation requires appropriate numerical schemes. This is in sharp contrast to the notion of the S-matrix which is identified with a linear operator mapping between infinite-dimensional function spaces \cite{Newton-ST} and leads to a rigorous mathematical theory of scattering \cite{reed-simon3,yafaev}. In 2016, we proposed a different notion of the transfer matrix in two and three dimensions, which shared this feature of the S-matrix \cite{pra-2016}. The main theoretical input leading to the introduction of this notion is the idea of expressing it in terms of the evolution operator for a non-unitary effective quantum system \cite{ap-2014}; hence the name ``dynamical formulation of stationary scattering.''

A more detailed examination of the developments reported in \cite{pra-2016} revealed an ambiguity whose resolution required a more careful treatment of the evanescent waves. In the present article, we propose a refinement of the conceptual framework offered in Ref.~\cite{pra-2016} that addresses this issue. It turns out that a proper implementation of this approach to stationary scattering entails the introduction of two different transfer matrices, which we call ``auxiliary and fundamental transfer matrices.'' Similarly to the S-matrix, these are linear operators mapping between infinite-dimensional function spaces. We have derived their basic properties and explored their utility in solving scattering problems. In particular, we have established and elucidated the origin of an implicit regularization property of the fundamental transfer matrix for delta-function potentials in two and three dimensions. We have also offered proper derivations of a couple of basic results on achieving perfect broadband omnidirectional invisibility (Theorem~2) and constructing potentials for which the first Born approximation yields the exact expression for the scattering amplitude (Theorem~3) in two dimensions. The results we have obtained for two dimensions admit straightforward extensions to three dimensions. In particular, we have obtained three-dimensional analogs of Theorems~2 and 3.

\section*{Appendix A: Proof of (\ref{thm1-114n})}

First, we introduce the function spaces
    \[\cS_\varsigma:=\{f\in\sF~|~f(p)=0~\for~p\leq\varsigma\},\quad\quad\quad \varsigma\in\R,\]
and prove a couple of lemmas.\\[6pt]
{\bf Lemma~1}: Let $\beta,\gamma\in\R$, $v$ be a short-range potential satisfying,
    \be
    \tilde v(x,\fK)=0~~\for~~\fK\leq \beta,
    \label{condi-4beta}
    \ee
and $g\in\sF$ be an arbitrary function such that
    \be
    g(p)=0~~\for~~p\leq\gamma,
    \label{gn-condi}
    \ee
i.e., for all $x\in\R$, $\tilde v(x,\cdot)\in\cS_\beta$ and $g\in\cS_\gamma$. Then $\widehat\sV(x)g\in\cS_{\beta+\gamma}$.\\[6pt]
{\bf Proof}: According to (\ref{v-def}) and (\ref{gn-condi}),
    \be
    \big(\widehat\sV(x)g\big)(p)=\frac{1}{2\pi}\int_{\gamma}^\infty dq\:
    \tilde v(x,p-q)g(q)=\frac{1}{2\pi}\int_{-\infty}^{p-\gamma} d\fK\:
    \tilde v(x,\fK)g(p-\fK).
    \label{lemma1-3}
    \ee
For $p\leq\beta+\gamma$, $p-\gamma\leq\beta$, and (\ref{condi-4beta}) implies that the integrand on the right-hand side of (\ref{lemma1-3}) vanishes. Therefore, $\widehat\sV(x)g\in\cS_{\beta+\gamma}$.~~$\square$\\[6pt]
{\bf Lemma~2}: Let $\alpha\in\R^+$, $\beta\in\R$, $k\in (0,\alpha]$, $\phi\in\sF_k$, $v$ be a short-range potential satisfying (\ref{condi-4beta}), $n\in\Z^+$, $x_1,x_2,\cdots,x_n\in\R$, $f_1,f_2,\cdots,f_n$ be functions of $p$, and
    \be
    \phi_n:=f_{n}(\hat p)\widehat{\sV}(x_n)f_{n-1}(\hat p)
    \widehat{\sV}(x_{n-1})f_{n-2}(\hat p)\widehat{\sV}(x_{n-2})\cdots f_{1}(\hat p)
    \widehat{\sV}(x_1)\phi.
    \label{phi-def}
    \ee
Then, $\phi_n(p)=0$ for $p\leq n\beta-\alpha$, i.e.,
    \be
    \phi_n\in\cS_{n\beta-\alpha}.
    \label{thm1}
    \ee
{\bf Proof} (by induction on $n$): Because $\phi\in\sF_k$ and $-\alpha\leq -k$, $\phi(p)=0$ for $p\leq -k$. This implies $\phi(p)=0$ for $p\leq-\alpha$. Hence $\phi\in\cS_{-\alpha}$. In view of this observation and (\ref{condi-4beta}), which means $\tilde v(x,.)\in\cS_\beta$, we can use Lemma~1 with $g=\phi$ and $\gamma=-\alpha$ to conclude that $\widehat\sV(x)\phi\in\cS_{\beta-\alpha}$. This together with the identity,
    \[\phi_1(p)=\big(f_{1}(\hat p)\widehat{\sV}(x_1)\phi\big)(p)=f_{1}(p)\big(\widehat{\sV}(x_1)\phi\big)(p),\]
imply $\phi_1\in\cS_{\beta-\alpha}$. Therefore, (\ref{thm1}) holds for $n=1$. Next, we suppose that there is a positive integer $m$ such that (\ref{thm1}) holds for $n=m$, and prove it for $n=m+1$. Clearly $\phi_{m+1}=f_{m+1}(\hat p)\widehat{\sV}(x_{m+1})\phi_m$. According to (\ref{condi-4beta}), $\tilde v(x,\cdot)\in\cS_\beta$, and induction hypothesis states that $\phi_m\in\cS_{m\beta-\alpha}$. By virtue of Lemma~1, these imply $\widehat{\sV}(x_{m+1})\phi_m\in\cS_{(m+1)\beta-\alpha}$. Combining this relation with
$\phi_{m+1}(p)=f_{m+1}(p)\Big(\widehat{\sV}(x_{m+1})\phi_m\Big)(p)$, we are led to $\phi_{m+1}\in\cS_{(m+1)\beta-\alpha}$. This proves (\ref{thm1}) for $n=m+1$.~~$\square$

Now, suppose that (\ref{condi-4beta}) holds for some $\beta>0$, $k\in(0,\alpha]$, and $\xi\in\sF$ be  arbitrary. Then applying Lemma~2 for $\phi:=\widehat\Pi_k\xi$ and $f_n(p)=1$, we find
    \be
    \Big(\widehat{\sV}(x_n)f_{n-1}(\hat p)
    \widehat{\sV}(x_{n-1})f_{n-2}(\hat p)\widehat{\sV}(x_{n-2})\cdots f_{1}(\hat p)
    \widehat{\sV}(x_1)\widehat\Pi_k\xi\Big)(p)=0~~\for~~p\leq n\beta-\alpha.
    \nn
    \ee
This relation together with the fact that $(\Pi_k\xi)(p)=0$ for $p>k$ imply
    \be
    \widehat{\sV}(x_n)f_{n-1}(\hat p)
    \widehat{\sV}(x_{n-1})f_{n-2}(\hat p)\widehat{\sV}(x_{n-2})\cdots f_{1}(\hat p)
    \widehat{\sV}(x_1)\widehat\Pi_k\xi=0~~\for~~n\geq\frac{k+\alpha}{\beta}.
    \nn
    \ee
This establishes (\ref{thm1-114n}), because $\xi$ is an arbitrary element of $\sF$.

\section*{Appendix B: Proofs of Theorems 4 and 5}

Let $\bfe_1$ be a unit vector belonging to the $x$-$y$ plane, and $\bfe_2:=\bfe_z\times\bfe_1$, where $\times$ stands for the cross product of vectors belonging to $\R^3$. Then, $\{\bfe_1,\bfe_2\}$ forms an orthonormal basis of the $x$-$y$ plane, and we can use it to introduce a Cartesian coordinate system for this plane. Let $x'$ and $y'$ denote the coordinates along the axes defined by $\bfe_1$ and $\bfe_2$, respectively. Clearly there is a rotation angle $\varphi\in[0,2\pi)$ such that
    \begin{align}
    &x'=\cos\varphi\,x-\sin\varphi\,y,
    &&y'=\sin\varphi\,x+\cos\varphi\, y.\nn
    \end{align}
Similarly, if $\vec p=p_x\,\bfe_x+p_y\,\bfe_y$ for some $(p_x,p_y)\in\R^2$, we
can express the components of  $\vec p$ in the coordinate frame $\{\bfe_1,\bfe_2\}$ as
    \begin{align}
    &p_1=\cos\varphi\,p_x-\sin\varphi\,p_y,
    &&p_2=\sin\varphi\,p_x+\cos\varphi\, p_y.\nn
    \end{align}

Next, for each $\varphi\in[0,2\pi)$ and $\varsigma\in\R$, we use $\sR_{\varphi,\varsigma}$ to denote the subset of the $x$-$y$ plane defined by $x'\leq\varsigma$, i.e.,
    \[ \sR_{\varphi,\varsigma}:=\big\{(x,y)\in\R^2~\big|~\cos\varphi\,x-\sin\varphi\,y\leq\varsigma~\big\},\]
and
    \[\cS_{\varphi,\varsigma}:=\left\{f\in\sF~|~f(\vec p)=0~\for~\vec p\in
    \sR_{\varphi,\varsigma}\right\},\]
where $\sF$ denotes the set of functions of $\vec p$.
We are now in a position to state three-dimensional analogs of Lemmas~1 and~2 of Appendix~A and Theorem~1.\\[6pt]
{\bf Lemma~3}: Let $\varphi\in[0,2\pi)$, $\beta,\gamma\in\R$, $v:\R^3\to\C$ be a short-range potential such that for all $z\in\R$, $\tilde{\tilde v}(\cdot,z)\in\cS_{\varphi,\beta}$, and $g\in\cS_{\varphi,\gamma}$. Then $\widehat\sV(z)g\in\cS_{\varphi,\beta+\gamma}$.\\[6pt]
{\bf Proof}: Using $g\in\cS_{\varphi,\gamma}$ in (\ref{eq-149}), we have
    \bea
    \big(\widehat\sV(z)g\big)(\vec p)&=&\frac{1}{4\pi^2}\int_{\R^2\setminus\sR_{\varphi,\gamma}}\!\!d^2\vec q\: \tilde{\tilde v}(\vec p-\vec q)g(\vec q)=
    \frac{1}{4\pi^2}\int_{-\infty}^\infty\!\! dq_2\int_\gamma^\infty\!\! dq_1
    \tilde{\tilde v}(p_1-q_1,p_2-q_2,z)g(q_1,q_2)\nn\\
    &=&
    \frac{1}{4\pi^2}\int_{-\infty}^\infty\!\! dq_2\int_{-\infty}^{p_1-\gamma}\!\! d\fK_1
    \tilde{\tilde v}(\fK_1,p_2-q_2,z)g(q_1-\fK_1,q_2).\nn
    \eea
The integrand on the right-hand side of this equation vanishes for $p_1\leq\beta+\gamma$, because $\tilde{\tilde v}(\cdot,z)\in\cS_{\varphi,\beta}$ and $\fK_1\leq p_1-\gamma\leq\beta$.~~$\square$\\[6pt]
{\bf Lemma~4}: Let $\varphi\in[0,2\pi)$,  $\alpha\in\R^+$, $\beta\in\R$, $k\in (0,\alpha]$, $\phi\in\sF_k$, $v:\R^3\to\C$ be a short-range potential such that for all $z\in\R$, $\tilde{\tilde v}(\cdot,z)\in\cS_{\varphi,\beta}$, $n\in\Z^+$, $z_1,z_2,\cdots,z_n\in\R$, $f_1,f_2,\cdots,f_n\in\cF$, and
    \be
    \phi_n:=f_{n}(\,\widehat{\vec p}\,)\widehat{\sV}(z_n)f_{n-1}(\,\widehat{\vec p}\,)
    \widehat{\sV}(z_{n-1})f_{n-2}(\,\widehat{\vec p}\,)\widehat{\sV}(z_{n-2})\cdots f_{1}(\,\widehat{\vec p}\,)
    \widehat{\sV}(z_1)\phi.
    \nn
    \ee
Then, $\phi_n\in\cS_{\varphi,n\beta-\alpha}$.\\[6pt]
{\bf Proof}: This follows from the inductive argument used in the proof of Lemma~2 and the fact that $k\in (0,\alpha]$ and $\phi\in\sF_k$ imply $\phi\in\cS_{\varphi,-\alpha}$.~~$\square$\vspace{6pt}

The following three-dimensional analog of Theorem~1 is an immediate consequence of Lemmas~3 and~4.\\[6pt]
{\bf Theorem~6}: Let $\varphi\in[0,2\pi)$, $\alpha,\beta\in\R^+$, $k\in(0,\alpha]$, and $v:\R^3\to\C$ be a short-range potential such that for all $z\in\R$, $\tilde{\tilde v}(\vec\fK,z)=0$ for $\fK_1\leq\beta$. Then, $\widehat{\breve\bM}=\widehat\bI$ for $\beta\geq 2\alpha$, and
    \be
    \widehat{\breve\bM}=\widehat\bI+\sum_{n=1}^{\lceil 2\alpha/\beta-1\rceil} (-i)^n\!\!
            \int_{z_0}^z \!\!dz_n\int_{z_0}^{z_n} \!\!dz_{n-1}
            \cdots\int_{z_0}^{z_2} \!\!dx_1\,\widehat\bPi_k
            \widehat{\breve\bcH}(z_n)\widehat{\breve\bcH}(z_{n-1})\cdots
            \widehat{\breve\bcH}(z_1)\widehat\bPi_k,
            \label{Thm-01-3D}
        \ee
for $0<\beta<2\alpha$.\\[6pt]
Theorems~4 and 5 follow as corollaries of Theorem~6; their hypotheses imply that of Theorem~6 with $\beta=2\alpha$ and $\beta=\alpha$, respectively. This shows that
    \begin{itemize}
    \item[-] if $\tilde{\tilde v}(\vec\fK,z)=0$ for $\fK_1\leq2\alpha$, $\widehat{\breve\bM}=\widehat\bI$ and the potential is omnidirectionally invisible for $k\leq\alpha$;
    \item[-] if $\tilde{\tilde v}(\vec\fK,z)=0$ for $\fK_1\leq \alpha$, we have $\beta=\alpha$, $\lceil 2\alpha/\beta-1\rceil=1$, and only the first two terms on the right-hand side of (\ref{Thm-01-3D}) survive for $k\leq \alpha$.
    \end{itemize}
In the latter case, we have a closed-form expression for $\widehat{\breve\bM}$ which similarly to two dimensions allows for the determination of $\breve B^{l/r}_-$, $\breve A^{l/r}_+$, and $\ff^{l/r}$. Specifically, we obtain $\breve B^{l/r}_-$ and $\breve A^{l/r}_+$ by multiplying the right-hand sides of (\ref{thm3-4}) and (\ref{thm3-5}) by $2\pi$ and replacing $p_0$ with $\vec p_0$. Substituting these in (\ref{FL-3d}) and (\ref{FR-3d}) and making use of (\ref{ff=ff-3D}), we arrive at
    \be
    \ff(\vartheta,\varphi)=-\frac{\tilde{\tilde{\tilde v}}\big(\bk(\vartheta,\varphi)-\bk_0\big)}{4\pi},
    \label{app-B-f}
    \ee
where $\tilde{\tilde{\tilde v}}$ stands for the three-dimensional Fourier transform of $v$, i.e., $\tilde{\tilde{\tilde v}}(\bfK)=\int_{\R^3} d^3\bfr\:e^{-i\bfK\cdot\bfr}v(\bfr)$, and $\bk$ and $\bk_0$ are respectively the scattered and incident wave vectors, i.e.,
    \begin{align*}
    &\bk(\vartheta,\varphi)=k(\sin\vartheta\cos\varphi,\sin\vartheta\sin\varphi,\cos\vartheta),
&&\bk_0=k(\sin\vartheta_0\cos\varphi_0,\sin\vartheta_0\sin\varphi_0,\cos\vartheta_0).
    \end{align*}
This completes the proof of Theorem~5, because (\ref{app-B-f}), which provides an exact expression for the scattering amplitude of the potential, coincides with the outcome of applying the first Born approximation \cite[\S 7.2]{sakurai}.

\section*{Acknowledgements}
This work has been supported by the Scientific and Technological Research Council of Turkey (T\"UB\.{I}TAK) in the framework of the project 120F061 and by Turkish Academy of Sciences (T\"UBA).

\ed